\documentclass[%
 reprint,
usenatbib,
 amsmath,amssymb,
 aps,
 nofootinbib, 
twocolumn]{aastex631}

\usepackage{rotating}
\usepackage{graphicx}
\usepackage{dcolumn}
\usepackage{tabularx}
\usepackage{bm}
\usepackage{hyperref}

\usepackage{booktabs, tabularx, makecell, multirow, xspace}
\usepackage{graphicx}	
\usepackage{amsmath}	
\usepackage{amssymb}	

\usepackage{physics}   
\usepackage{comment}   
\usepackage{slashed}

\newcommand{\K}{\,{\rm K}}

\renewcommand{\arraystretch}{1.8}
\newcolumntype{C}[1]{>{\centering\let\newline\\\arraybackslash\hspace{0pt}}m{\#1}}

\def\aap{A\&A}

\def\apjl{ApJ}

\def\apjs{ApJS}

\newcommand{\Gaia}{$Gaia$\ }
\newcommand{\Cloudy}{{\texttt{Cloudy}}}

\newcommand{\sref}[1]{Section~\ref{#1}}

\newcommand{\fref}[1]{Figure~\ref{#1}}
\newcommand{\eref}[1]{Eqn.~\eqref{#1}}

\newcommand\blfootnote[1]{%
  \begingroup
  \renewcommand\thefootnote{}\footnote{#1}%
  \addtocounter{footnote}{-1}%
  \endgroup
}

\begin{document}


\title{Electromagnetic Signatures of Mirror Stars}

\author[0009-0006-1587-719X]{Isabella Armstrong*}
\affiliation{Department of Astronomy and Astrophysics, University of Toronto, Toronto, Ontario M5S 1A7, Canada}

\author[0009-0000-4150-4625]{Berkin Gurbuz*}
\affiliation{Department of Physics, University of Toronto, Toronto, Ontario M5S 1A7, Canada}

\author[0000-0003-0263-6195]{David Curtin}
\affiliation{Department of Physics, University of Toronto, Toronto, Ontario M5S 1A7, Canada}

\author[0000-0001-9732-2281]{Christopher Matzner}
\affiliation{Department of Astronomy and Astrophysics, University of Toronto, Toronto, Ontario M5S 1A7, Canada}

\begin{abstract}
Mirror Stars are a generic prediction of dissipative dark matter models, including minimal atomic dark matter and twin baryons in the Mirror Twin Higgs. 
Mirror Stars can capture regular matter from the interstellar medium through extremely suppressed kinetic mixing interactions between the regular and the dark photon. This accumulated ``nugget'' will draw heat from the mirror star core and emit highly characteristic X-ray and optical signals.
In this work, we devise a  general parameterization of mirror star nugget properties that is independent of the unknown details of mirror star stellar physics, and use the \texttt{\texttt{Cloudy}} spectral synthesis code to
obtain realistic and comprehensive predictions for the thermal emissions from optically thin mirror star nuggets. 
We find that mirror star nuggets populate an extremely well-defined and narrow region of the HR diagram that only partially overlaps with the white dwarf population. 
Our detailed spectral predictions, which we make publicly available, allow us to demonstrate that optically thin nuggets can be clearly distinguished from white dwarf stars by their continuum spectrum shape, and from planetary nebulae and other optically thin standard sources by their highly exotic emission line ratios. Our work will enable realistic mirror star telescope searches, which may reveal the detailed nature of dark matter. 
\end{abstract}

\section{Introduction}
\label{sec:intro}
\blfootnote{\emph{First authors are indicated with *.}}
While the existence of dark matter (DM) is essentially undisputed, very little is known about its detailed nature or interactions.  
The prevailing paradigm is cold dark matter (CDM). Assumed to be collisionless on astrophysical scales, and typically imagined as a single type of new subatomic particle like the WIMP (Weakly Interacting Massive Particle)~\citep{Battaglieri:2017aum}, CDM
provides an excellent description of the large-scale structure of our universe~\citep{Planck:2018vyg}, and accounts for the invisible mass component inferred through its gravitational influence in galaxies and galaxy clusters. 
While CDM has been a successful benchmark, and seems on the surface to be a pleasingly minimal model, 
there is ample  motivation to go beyond this simple framework. 

Empirically, discrepancies between CDM simulations and observational data at dwarf galaxy scales~\citep{Bullock:2017xww,Governato:2009bg,Garrison_Kimmel_2019, Relatores_2019}
suggest the possibility of DM self-interactions.
Theoretically, it is highly plausible that DM is part of a more complex dark sector that includes multiple particle species and forces, which could account for the small-scale discrepancies~\citep{Tulin:2013teo}.
From a bottom-up point of view, this is motivated by analogy to the highly non-minimal structure of the Standard Model (SM) of particle physics. 
This possibility is studied by using simplified benchmark models like Atomic Dark Matter (aDM)~\citep{Kaplan:2009de}, which postulate the existence of a dark analogue of electromagnetism mediated by a massless (and to us invisible) dark photon, under which a heavier and a lighter dark matter state (dubbed dark proton and dark electron) are oppositely charged.

Importantly, many theories of beyond-SM (BSM) physics that address various long-standing mysteries in fundamental physics predict the existence of exactly such a dark sector.
The most relevant examples are models of Neutral Naturalness such as the Mirror Twin Higgs \citep[MTH:][]{Chacko:2005pe, Chacko:2016hvu, Craig:2016lyx}, which solve the electroweak hierarchy problem in a different manner from canonical frameworks like supersymmetry~\citep{Martin:1997ns}, making them compatible with null results from the Large Hadron Collider: see reviews by \citet{Batell:2022pzc} and \citet{Craig:2023lyt}.
The MTH includes a complex dark sector related to the SM by a discrete symmetry, thereby predicting the existence of twin protons, twin neutrons, and twin electrons interacting via twin versions of the SM forces, essentially implementing aDM with nuclear interactions.
This discrete symmetry also makes it highly plausible that aDM, or something like it, makes up a \emph{fraction} of the total DM abundance, since whatever unknown mechanism of baryogenesis produces SM baryons in the early universe, which make up about 1/6 of the total matter budget, would be expected to have a dark-baryon equivalent as well (see ~\cite{Farina:2015uea, Earl:2019wjw, Alonso-Alvarez:2023bat} for concrete theoretical proposals). Such aDM sub-components are compatible with overall DM self-interaction constraints~\citep{Randall:2008ppe}, even if they behave very differently from CDM.

aDM is not only self-interacting: it can form dark atomic bound states and is \emph{dissipative}, meaning it can cool by emitting dark photons and collapse to form structure.
While aDM would therefore leave its imprint in early-universe cosmology~\citep{Cyr-Racine:2013fsa,
Bansal:2022qbi,
Bansal:2021dfh,
Zu:2023rmc} and galactic dynamics~\citep{Fan:2013yva, Ghalsasi:2017jna, Roy:2023zar, Gemmell:2023trd},\footnote{See also 
\cite{Fan:2013tia,McCullough:2013jma,Randall:2014kta,Schutz:2017tfp, Buch:2018qdr,Ryan:2021dis, Gurian:2021qhk, Ryan:2021tgw, Foot:2013lxa,  Foot:2014uba, Foot:2015mqa,Chashchina:2016wle, Foot:2017dgx, Foot:2018dhy, Foot:2016wvj, Foot:2013vna}.} one of its most spectacular predictions is the existence of \emph{Mirror Stars}.

Mirror Stars are, in perfect analogy to regular stars, collapsed clumps of aDM that radiate their internal energy away as invisible dark photons.
These mirror stars can capture regular matter from the interstellar medium due a tiny mixing between dark and SM photons, which quickly settles into the center, in analogy to the hypothetical capture of DM by the Sun \citep{1985ApJ...296..679P,1987ApJ...321..571G}. 
It was recently established~\citep{Curtin:2019ngc, Curtin:2019lhm} that the captured SM matter 
is heated by the core and gives off highly characteristic optical and X-ray emissions.
While faint, emissions by these captured \emph{nuggets} are detectable in our stellar neighborhood, representing a unique discovery opportunity for atomic dark matter using telescope observations.

Previous work demonstrated how mirror stars could be searched for in public \Gaia data~\citep{Howe:2021neq}, since the nuggets' high temperature $(T \sim 10^4 K)$ but extreme faintness distinguishes mirror stars from standard astrophysical point sources. However, this analysis used highly simplified bremsstrahlung-only calculations for the emissions of the captured nuggets. 
While this gives a passable estimate of the overall magnitude, it does not represent realistic emission spectra.
Improved estimates are needed to obtain quantitatively correct predictions for the colour of mirror stars in a Hertzsprung-Russell diagram and conduct realistic \Gaia searches. 
Detailed emission calculations will also enable searches for mirror stars that rely purely on their unique spectral characteristics, without requiring  parallax measurements.

In this paper, we present realistic predictions for the electromagnetic emission spectra of mirror stars (via their captured SM nuggets).
Astrophysically, optically thin nuggets behave much like compressed planetary nebula, and just like planetary nebulae, their emissions are dominated by atomic processes, which determine both their temperatures and their spectral properties.
We therefore use the nebular synthesis code \texttt{\texttt{Cloudy}} \citep{ferland1998cloudy}, a sophisticated platform for computing the thermal,  ionization, and dynamical equilibria of interstellar clouds, along with their emitted spectra. \texttt{Cloudy} can be adapted to solve for the equilibrium configuration of SM gas that sits in the gravitational well and is heated by the mirror star core, making it ideal for the task at hand.
The optical mirror star signal can be parameterized in a very general way in terms of the heating rate from the core to the nugget; the total nugget mass; and the core density of the mirror star. This allows us to capture the full range of mirror star emission signals, without having to specify any details of mirror star stellar physics, lifetime, capture processes, etc. 

We make our \Cloudy~input cards and full results publicly available.\footnote{\url{https://github.com/davidrcurtin/mirror_star_emissions}} This will enable a wide range of new optical searches for mirror stars. 

This paper is structured as follows. We briefly review the physics of mirror stars, how they capture SM matter, and the resulting electromagnetic emissions in \sref{sec:review}. We also discuss how to generally parameterize the mirror star signal without specifying all the details of the mirror star or its constituent dark sector or aDM particles. 
In \sref{s.modelingnugget}, we discuss how to use the \texttt{\texttt{Cloudy}} code to model optically thin nuggets captured by mirror stars, and compute their emissions. 
The physical properties of our nugget solutions are also discussed. 
Our predictions for mirror star optical signatures across the possible parameter space are presented in \sref{s.results}.
\sref{s.HR} presents mirror star emissions in a Hertzsprung-Russell diagram to establish a robust signal region for mirror star with optically thin SM nuggets, and \sref{s.spectra} presents the resulting emissions spectra and how they can be distingushed from standard astrophysical sources like planetary nebulae or white dwarfs.
We conclude by discussing the implications of our work, including direct application to a variety of possible mirror star telescope searches, in \sref{s.conclusion}.

\vspace{5mm}

\section{Mirror Star Review \label{sec:review}}

Since atomic dark matter is dissipative and self-interacting, it will cool and collapse and form compact objects, in direct analogy to star formation for SM baryons. 
If the aDM additionally has dark nuclear interactions, as it does in models like the Twin Higgs, then these mirror stars 
may undergo dark fusion processes in their cores and generically shine dark photons for astrophysically long periods of time. If the aDM is minimal and does not feature dark nuclear reactions, {or those reactions exist but do not ignite,  a mirror star will still 
evolve for Kelvin-Helmholtz timescales,  {before either forming a cooling, degenerate object or (if its mass exceeds the dark Chandrasekhar limit) collapsing to a dark neutron star or black hole. In any case, low-mass mirror stars, which probably dominate the population as they do for normal stars, are likely to collect and heat SM material for $\sim10^{8-10}$ years or more, for a broad range of aDM parameters.

The range of possible properties of these mirror stars will be determined by the microphysics parameters of the aDM: its constituent masses, and its dark-electromagnetic coupling. The actual distribution of mirror stars is additionally determined by the detailed cosmological and astrophysical evolution of the atomic dark matter density perturbations from the early universe through to galaxy formation and local star formation today. Predicting the former from the microphysics would be challenging but possible; predicting the latter with a degree of reliability is as difficult as predicting the shape of the Milky Way and the stellar mass function from the SM Lagrangian. 
Fortunately, we can understand the general properties of mirror stars and their possible discovery signals without relying on such detailed model-specific predictions.

We focus on electromagnetic signals of mirror stars, but microlensing surveys will also be highly sensitive in the near future, see 
\citet{Winch:2020cju}.
Furthermore, mirror stellar relics are  expected to give rise to distinctive gravitational wave signals from merger events, including mirror white dwarfs~\citep{Ryan:2022hku}, mirror neutron stars~\citep{Hippert:2021fch, Hippert:2022snq} and black holes with unusual masses sourced by aDM collapse~\citep{Pollack:2014rja, Shandera:2018xkn, Singh:2020wiq, Gurian:2022nbx, Fernandez:2022zmc}.

\subsection{Capture and Heating of SM Matter}

We now briefly summarize the most important points of the analyses by~\cite{Curtin:2019ngc, Curtin:2019lhm}, which demonstrated the existence of observable optical and X-ray signatures of mirror stars.\footnote{See \cite{Foot:1999hm, Foot:2000vy, Foot:2004pa, Foot:2014mia} for earlier work discussing the possibility of mirror stars.}

If aDM or something like it exists, then it is highly likely that the massless dark photon \emph{kinetically mixes} with the SM photon. This is expressed as the Lagrangian operator $\mathcal{L} \supset \frac{1}{2}\epsilon F_{\mu \nu} F_D^{\mu \nu}$, with $F, F_D$ being the field strength of the SM and dark photon respectively. While the kinetic mixing $\epsilon$ can formally be of any size, it is generally expected that this parameter is \emph{dynamically generated} by quantum mechanical loop effects connecting the SM and dark sector at some order. Estimates of quantum-gravity contributions motivate $\epsilon \sim 10^{-14} - 10^{-13}$~\citep{Gherghetta:2019coi}, while the accidental symmetries of the MTH with a minimal UV completion suggest $\epsilon \sim 10^{-13} - 10^{-10}$~\citep{Koren:2019iuv}. Such tiny mixings have negligible cosmological and astrophysical effects, except for the fact that they enable capture of SM matter in mirror stars \citep[and vice versa, see e.g.,][]{Curtin:2020tkm}, and allow for direct detection of ambient aDM~\citep{Chacko:2021vin, SENSEI:2020dpa, SuperCDMS:2022kse}.

The kinetic mixing interaction results in the capture of interstellar SM matter as the mirror star traverses our Milky Way. For mirror stars with similar properties to regular stars, this is already highly efficient for $\epsilon \gtrsim 10^{-12}$, resulting in asteroid-mass-scale amounts of interstellar gas accumulating in the mirror star and forming a small \emph{nugget} in the center. 
The SM gas is then heated through $\epsilon$-suppressed electromagnetic interactions with the hot mirror star core.
The heating rate per SM atom is (in natural particle physics units)
\begin{equation}
    \label{e.Pheating}
    \Gamma_n^{coll} \approx n' \frac{2 \pi \epsilon^2 \alpha \alpha' Z^2 {Z'}^2}{m} \Bigg \langle \frac{1}{v_{rel}} \left( \log \frac{8 \mu^2 v_{rel}^2}{(1/a_0)^2} - 1 \right) \Bigg \rangle \ .
\end{equation}
This expression applies for collisions involving a single SM (aDM) gas species with nuclear mass $m$ ($m'$), atomic number $Z$ ($Z'$) and atomic number density $n$ $(n')$, where the SM gas is assumed to be mostly non-ionized (resulting in an approximate screening length\footnote{A more precise treatment would replace $a_0$ by the Debye length of the SM plasma if it is highly ionized, but as the exact screening length has only a very minor effect on the signal, and since the overwhelming majority of the optical nugget signal is generated at temperatures with relatively low ionization fractions, \eref{e.Pheating} is an acceptable approximation.} of the Bohr radius $a_0$), and
much colder than the mirror star core temperature $T_{{\rm core}}$.
For nuggets and mirror star cores with multiple (dark) atomic species, the above expression is generalized to sums of SM-dark collision pairs.

These interactions heat up the captured nugget to temperatures of order  $T_{{\rm nugget}} \sim 10^{4}$\,K, at which point atomic and other cooling processes become very efficient and easily radiate away the energy siphoned from the mirror star core through $\epsilon^2$-suppressed collisions.
Since the heating rate is independent of the nugget temperature, the total luminosity can be simply found by multiplying \eref{e.Pheating} by the number of constituent SM atoms for each species. For mirror stars that are similar to regular stars, this can be as bright as a faint white dwarf, or orders of magnitudes dimmer, depending on $\epsilon$ and the mirror star lifetime (which proportionally determines the accumulated mass).

While the nugget luminosity is fairly straightforward to determine, and could clearly be detectable in our stellar neighborhood with various telescope surveys, more careful calculations are required to predict the detailed emissions spectrum. 
The nuggets sit in an external gravitational potential, which can be approximated by a constant mass density given by the central mirror star core density $\rho_{{\rm core}}$.
This assumes that the nugget is much smaller than the mirror star core, but the resulting scale height 
\begin{equation}
    \label{e.hnugget}
    h_{{\rm nugget}} \sim \sqrt{\frac{k T_{{\rm nugget}}}{G \rho_{{\rm core}} \bar m}}
\end{equation} 
is $\mathcal{O}(10^3\mathrm{km})$ for $\rho_{{\rm core}} = \rho_{{\rm core}, \odot} = 160\,$g/cm$^3$, our sun's core density.  This assumption is therefore very likely to be satisfied.
There is only modest dependence on the nugget composition  through the average nuclear constituent mass $\bar m$.
With nugget size and temperature only weakly depending on the total nugget mass, it is therefore clear that nuggets below some mass threshold will be diffuse enough to be optically thin, and cool via volume emission from atomic and free-free processes, while heavier nuggets are optically thick, and cool via surface black-body emission, analogous to tiny regular stars. In this work, we will focus on optically thin nuggets.

One of the potentially most exciting signals of mirror stars are the X-ray emissions of the captured nugget. The presence of regular matter and aDM in the same space allows for \emph{X-ray conversion}, where thermal dark photons from the mirror star core elastically scatter off SM electrons and nuclei with $\epsilon^2$ suppressed rates, to be re-emitted as a SM photon of the same energy. 
In other words, the captured nugget shines with a faint, approximately black body X-ray spectrum, given by the mirror star core temperature $T_{{\rm core}}$, in addition to its own thermal emissions set by $T_{{\rm nugget}}$. The conjunction of these two unusual signals is one of the most robust smoking guns of mirror stars~\citep{Curtin:2019ngc, Curtin:2019lhm}.

There are subtleties in computing the X-ray signal, since the nugget is generally optically thick at such high frequencies, so only X-rays emitted in a shell near the nugget surface escape with some modulation of their spectrum. However, 
we will focus on the regime where the amount of energy transferred to the nugget via X-ray conversion relative to the collisional heating of \eref{e.Pheating} is small. This ratio is given by
\begin{equation}
    \label{e.PconvoverPcoll}
    \frac{\Gamma_n^{conv}}{\Gamma_n^{coll}} \sim 10^{-5} \left( \frac{\rho_{{\rm core}, \odot}}{\rho_{{\rm core}}}\right) \left( \frac{m_{H'}}{m_{H}}\right) \left( \frac{T_{{\rm core}}}{10^7 K}\right)^{9/2} \ .
\end{equation}
For core densities and temperatures somewhat similar to regular stars, as normalized above, it is indeed valid to ignore the contribution of X-ray conversion to the nugget heating for the purpose of computing its optical signature. However, the steep $T_{{\rm core}}$ dependence means that X-ray heating may dominate for aDM scenarios with much hotter mirror star interiors. We leave this interesting possibility for future investigation, and focus in this paper exclusively on carefully determining the thermal emissions of optically thin nuggets due to the collisional heating of \eref{e.Pheating}.

\subsection{Parameterization of Mirror Star Signal}

The properties and emissions of the captured nuggets depend on the properties of the mirror star, the aDM microphysics (masses and couplings), and the size of the photon kinetic mixing portal. Fortunately, a very compact subset of parameters is sufficient to uniquely determine the characteristics of the nugget, and we can vary them without specifying the underlying dark sector physics:
\begin{enumerate}
    \item \emph{Nugget mass} $M_{{\rm nugget}}$, with more massive nuggets extracting more energy from the mirror star core and being therefore more luminous. 
    \item \emph{Mirror star central core density} $\rho_{{\rm core}}$, which determines the gravitational potential in which the nugget settles into hydrostatic equilibrium.
    \item \emph{Heating rate} $\xi$ from the mirror star to the nugget, which determines the total luminosity. 
\end{enumerate}
The last parameter requires some explanation. 
First, we choose a convention where rather than summing over each constituent species (H, He, C, \ldots) in the captured gas of the nugget, we express the \emph{total} heating rate into the nugget from all types of collisions in terms of the local hydrogen number density:
\begin{equation}
    \label{e.Pheating2}
    \Gamma_H\ \  \approx \  
    \left( \frac{\rho_{{\rm core}}}{\rho_{{\rm core}, \odot}}\right) \ 
    \xi \ 
    (\mathrm{J/s}) \ .
\end{equation}
The total heating rate of the nugget per unit volume is therefore $\Gamma_H n_H$, which is compatible with how \Cloudy\  handles heating rates (all per-atom heating rates and non-hydrogen number densities are given in terms of $n_H$).
We  define the dimensionless heating rate parameter $\xi$, which is taken as an input parameter to  determine nugget properties, but it can be related to microphysical parameters, as well as the detailed nugget and mirror star core composition, using \eref{e.Pheating}:
\begin{eqnarray}
    \nonumber \xi &\approx& (0.45) \epsilon^2 \left(\frac{\alpha_D}{\alpha}\right)  \left( \frac{1.5 \times 10^7 K}{T_{{\rm core}}}\right)^{\frac{1}{2}}  \left( \frac{m_{H}}{m_{H'}} \right)^{\frac{1}{2}} 
    \left(\frac{L_{HH'}}{14}\right)
    \\
    && 
    \times \left[
    \sum_{i,i'}
    \frac{q_i \eta'_{i'} Z_i^2 {Z'_{i'}}^2}{A_i \sqrt{A'_{i'}}} \frac{L_{ii'}}{L_{HH'}}
    \right] \ .
\end{eqnarray}
where we adopt the convention of labeling the lightest dark nuclear state as ``dark hydrogen'' $H'$ (without necessarily assuming that its dark electric charge is 1). The dimensionless term in square brackets accounts for the composition of the nugget and the mirror star core beyond pure (dark) hydrogen, and is generally expected to be $\mathcal{O}(1)$.\footnote{The exact physical interpretation of a given heating rate parameter $\xi$ is therefore slightly dependent on the nugget composition, but since $\xi$ varies over many orders of magnitude due to the unknown tiny kinetic mixing factor $\epsilon^2$, the convenience of this choice justifies a slight convolution of SM and BSM parameters.} We parameterize the local nugget composition relative to the hydrogen number density
$n_i = q_i n_H$; $\eta'_{i'}$ is the mass fraction of dark nuclear component $i'$ in the core; we define generalized atomic numbers $A_i = m_i/m_H, A'_{i'} = m_{i'}/m_{H'}$; and the log factor is
\begin{equation}
    L_{ii'} \equiv \log\left[8 a_0^2 \mu_{ii'}^2 \frac{3 k T_{{\rm core}}}{m_{i'}} \right]
    - 1 \ .
\end{equation}
where $\mu_{ii'}$ is the reduced mass of $i-i'$ collisions.
This log factor only has very slight variation with aDM and mirror star parameters: $L_{HH'} \sim 14$ for core temperatures and dark hydrogen masses close to their SM counterparts, and the ratios $L_{ii'}/L_{HH'}$ in square brackets above will generally be very close to 1.
The usefulness of this parameterization is now clear: it is completely general with respect to nugget and mirror star composition, but for even vaguely SM-like aDM parameters and mirror star properties, we can roughly interpret $\xi \sim \epsilon^2$.

This parameterization covers both optically thin and optically thick nuggets, even as we focus on the former in this work. Though \eref{e.Pheating2} is obviously optimized to match the dependencies of collisional heating, our results can likely be applied, with some re-interpretation of $\xi$ to separate out the $\rho_{{\rm core}}$ dependence of \eref{e.PconvoverPcoll}, to determine the optical signal of nuggets dominantly heated by X-ray conversion as well (assuming most of the X-rays are absorbed by the nugget). To additionally describe the X-ray signal itself we would only have to minimally extend our parameterization to include separately the mirror star core temperature $T_{{\rm core}}$. We will actually make use of this flexibility to investigate the corona of the nugget in the next sections.

While our parameterization is sufficient to uniquely determine the signal of captured nuggets, it is useful to relate the nugget mass $M_{{\rm nugget}}$ to a rough \emph{mirror star lifetime} $\tau_{MS}$ required to actually collect this amount of SM interstellar material. For the kinds of small kinetic mixings we are most interested in, $\epsilon \sim 10^{-14} - 10^{-10}$, the fastest possible accumulation mechanism is capture of ISM material by collisions with the nugget itself.\footnote{Indeed, this capture process quickly becomes dominant early in the lifetime of the SM-like mirror stars studied in~\cite{Curtin:2019ngc}.} Assuming an approximate nugget size given by the scale height in \eref{e.hnugget}, geometric capture is valid for $M_{{\rm nugget}} \gtrsim (10^{10} g) (\rho_{{\rm core}, \odot}/\rho_{{\rm core}})$, which is smaller than any nugget size we consider in our analysis. Assuming an interstellar medium density of one hydrogen atom per $cm^3$ and a one-dimensional velocity dispersion of  $\sim$30\,km/s characteristic of our local stellar environment~\citep[][Table 1.2]{binney2011galactic}, the nugget mass accumulated from geometric capture during a mirror star lifetime of $\tau_{MS}$ 
is
\begin{equation}
    M_{{\rm nugget}} \sim \left(10^{16} g\right) \left(\frac{\rho_{{\rm core}, \odot}}{\rho_{{\rm core}}}\right)
    \left(
    \frac{\tau_{MS}}{10^9 \ \mathrm{years}}\right) \ .
\end{equation}
Our analysis will enable us to understand the full range of $M_{{\rm nugget}}$, $\rho_{{\rm core}}$ and $\xi$ that give rise to observable optically thin nuggets in mirror stars. This will correspond to core densities within two orders of magnitude of our sun, and nugget masses in the range $M_{{\rm nugget}} \sim 10^{14} - 10^{18} g$. Our results therefore correspond to potentially observable nuggets that would require mirror star ages in the million to billion year range, very reasonable possibilities for mirror star lifetimes both with and without dark nuclear fusion.

\section{Modeling the captured SM Nugget}
\label{s.modelingnugget}

The baryonic mirror star nuggets were modeled using \texttt{Cloudy} release C22~\citep{ferland20232017},
a physical and spectral synthesis code that solves for the ionization, thermal state, radiative transfer, and (optionally) hydrostatic equilibrium of diffuse nebulae, subject to the requirement of low to moderate optical depth.    
While mirror star nuggets are orders of magnitude more dense than planetary nebulae, they are often
sufficiently diffuse and optically thin for \texttt{Cloudy} to be able to find physical solutions to the nugget properties and emissions.
The calculation is exactly analogous to that of planetary nebulae,  with the external gravitational potential of stars in the nebulae replaced by the external gravitational potential of the mirror star core, and the heating from stars replaced by the constant heating rate of \eref{e.Pheating2}. 

\subsection{Numerical Integration with \texttt{Cloudy}}

The full details of how the mirror star nuggets are modeled are documented in our public results repository~[\href{https://github.com/davidrcurtin/mirror_star_emissions}{link}].
For our purposes, we work in a spherically symmetric geometry and impose the external gravitational potential generated by the constant background density $\rho_{{\rm core}}$.
\texttt{Cloudy}  starts with an initial hydrogen density at the center (which  determines the nugget mass $M_{{\rm nugget}}$)  and integrates outwards to self consistently compute the structure of the cloud as well as predicting its observed spectrum. 
\texttt{Cloudy} divides the cloud into a series of zones $k$ at radii $r_k$ and thickness $dr_k$, with each zone having constant temperature $T_k$, pressure $P_k$, density $\rho_k$, and ionization state $\chi_k$. 
Working outwards, the code self-consistently solves for hydrostatic equilibrium, continuity, radiative and heating balance, applying the heating rate of \eref{e.Pheating2} and from the CMB separately within each zone according to its density. We 
adopt
\texttt{Cloudy}'s default ISM abundances, which are an average for the warm and cold phases of the ISM \citep{ferland20232017}. The composition is predominantly 
H and He, 
with additional contributions from heavier elements \citep{Savage1996, Meyer1998, Snow2007, Mullman_1998}. 
Dust grains are ignored since, if they are captured, they will settle within the nugget.\footnote{They may subsequently be vaporized by heating from the mirror star, which raises questions beyond the scope of this paper.}

This yields self-consistent solutions for the nugget's structure (enclosed mass, temperature, pressure, and other physical properties as functions of radial distance $r$ from the mirror star center) as well as the emission spectrum differential in frequency.

The integration starts at the center and proceeds outwards until the stopping condition is met. This stopping condition deserves a careful discussion.

\subsection{Stopping condition and X-ray corona}
\label{s.stoppingcondition}

\begin{figure}
    \centering
    \includegraphics[width=0.4 \textwidth]{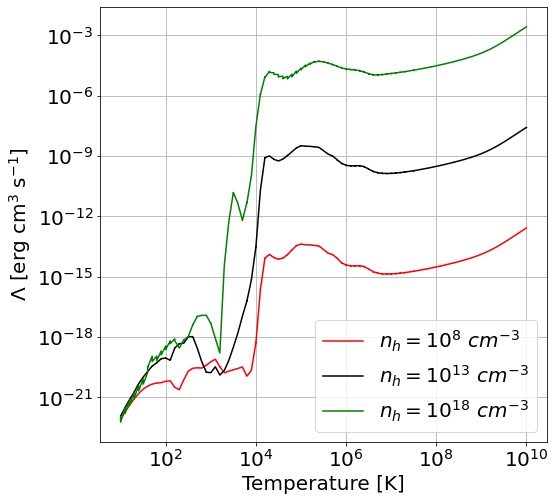}
    \caption{
    Cooling rate of baryonic gas at densities of $10^8$, $10^{13}$ and $10^{18}$ H atoms$/$cm$^3$ as a function of temperature, calculated with \Cloudy. 
    }
    \label{f.cooling}
\end{figure}

The total cooling rate must equal the heating rate in the nugget at equilibrium.  For optically thin nuggets, this equilibrium condition must be satisfied locally, i.e. in each zone separately.
The heating from the mirror star core is unusual, in that it is $\epsilon^2$-suppressed and can hence  be very small, but since it originates from a thermal bath at some presumably very high core temperature $T_{{\rm core}} \gg 10^4$\,K, this heating rate is essentially constant and independent of nugget temperature until $T_{{\rm nugget}} \sim T_{{\rm core}}$ (see \citealt{Curtin:2020tkm} for the detailed dependence on $T_{{\rm nugget}}/T_{{\rm core}}$). 

\fref{f.cooling} shows the cooling rate per hydrogen atom for ISM gas  obtained from a one-zone \texttt{Cloudy} calculation at different densities. We can imagine finding the equilibrium temperature of each zone  by finding the lowest temperature at which the cooling rate $\Lambda$ equals the total heating rate $\Gamma_H$.
Since the nugget density decreases monotonically with increasing $r$, leading to less efficient cooling in the outer zones, $T_{{\rm nugget}}(r)$ will increase with $r$.  However, the sharp rise in the cooling rate at $\sim10^4$\,K implies that most of the nugget will be at about this temperature for a wide range of heating rates. Note that
atomic cooling processes create a  a local maximum in the cooling rate at $T_{{\rm max}} \approx 8 \times 10^4 $\,K. If the heating rate exceeds this maximum, which it will for zones of sufficiently low density at some $r_{{\rm max}}$, then the  equilibrium temperature (assuming only standard-model cooling) automatically jumps to $\sim 10^9$ K.  If this is higher than the mirror star core temperature, then all zones with $r > r_{{\rm max}}$ will effectively thermalize with the mirror star core. 
However, since our simplified parameterization of the heating rate does not include 
the inverse of the heating process nor its $T_\mathrm{core}$ dependence,
we 
cannot use \texttt{Cloudy} to solve for 
zones that equilibrate with the mirror star.

We therefore terminate the \texttt{Cloudy} integration as soon as one of the zones exceeds $T_{{\rm max}}$. 
Physically, we can expect that for $r > r_{{\rm max}}$, the nugget gas would be extremely hot $(T \sim T_{{\rm core}})$ but also extremely diffuse, dominantly cooling by X-ray bremsstrahlung.
In other words, it forms a \emph{corona}. 

Realistically, this corona would roughly track the temperature profile and radial extent of the mirror star core. This is physically very interesting, and could reveal detailed information about mirror stellar physics if it could be observed.
To estimate the maximum possible impact of this corona on the electromagnetic nugget signal, we approximate it by an isothermal solution of hydrostatic equilibrium with $T = T_{{\rm core}} \gg 10^4 $\,K for $r > r_{{\rm max}}$ in the gravitational potential of constant $\rho_{{\rm core}}$ for arbitrarily high $r$. We evaluate this for a variety of mirror star core temperatures that are high enough for the corona to be completely ionized, and we match the corona pressure at $r = r_{{\rm max}}$ to the outermost zone solved by \texttt{\texttt{Cloudy}}. 
This analytical corona solution adds to the nugget mass and emission spectrum, with the latter contribution given by the one-zone \texttt{Cloudy} simulations we used to obtain the cooling curves in \fref{f.cooling}. 
However, ultimately this corona contribution to both nugget mass and luminosity is completely negligible by many orders of magnitude.

We can hence conclude that the thermal signal and total mass $M_{{\rm nugget}}$ of the nugget for a given $\rho_{{\rm core}}, \xi$ are well determined by the $T_{{\rm core}}$-independent \texttt{Cloudy} solution terminating at $r = r_{{\rm max}}$, where $T = T_{{\rm max}} = 8 \times 10^4$\,K. We therefore ignore the corona of the optically thin nuggets we study from here on.\footnote{If we wanted to also determine the observable X-ray signal, direct compton conversion, see \eref{e.PconvoverPcoll}, would completely dominate the contribution of the corona as well.}

\begin{figure}
    \centering
    \begin{tabular}{r}
    \includegraphics[height=4.5cm]{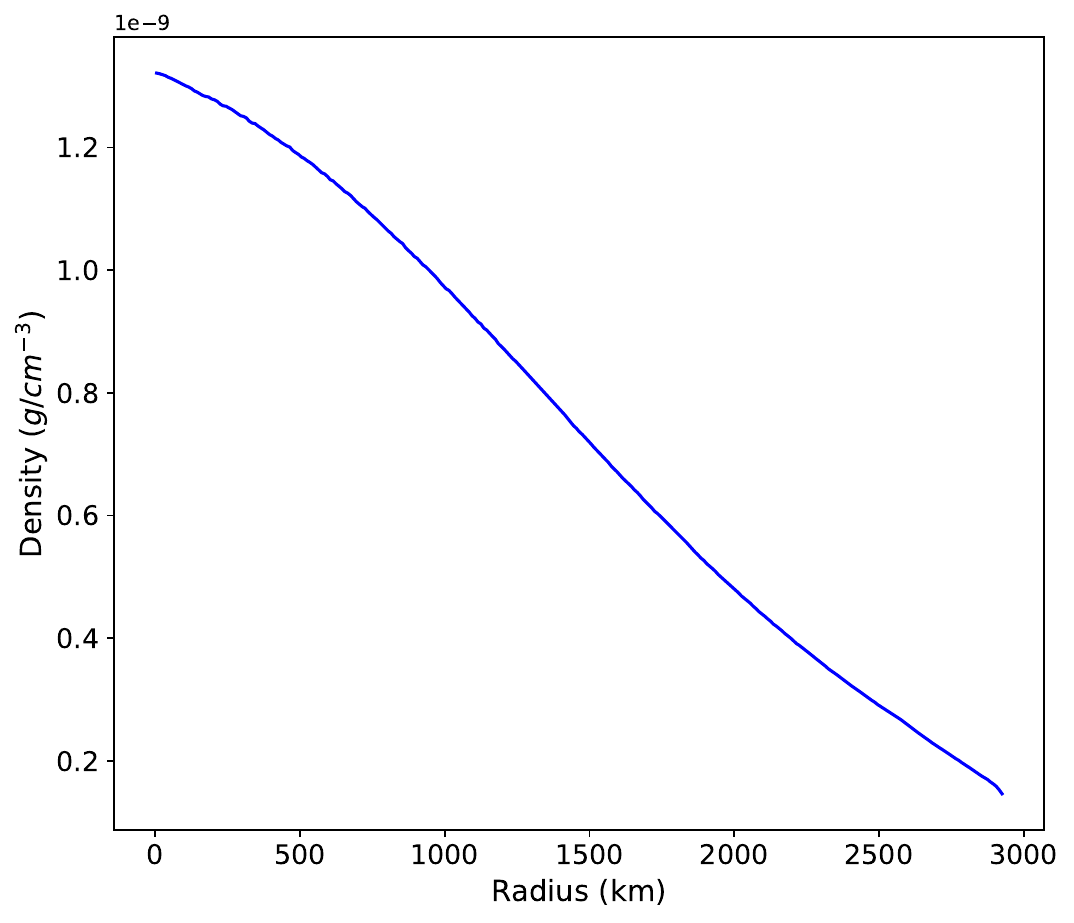}
    \\
    \includegraphics[height=4.42cm]{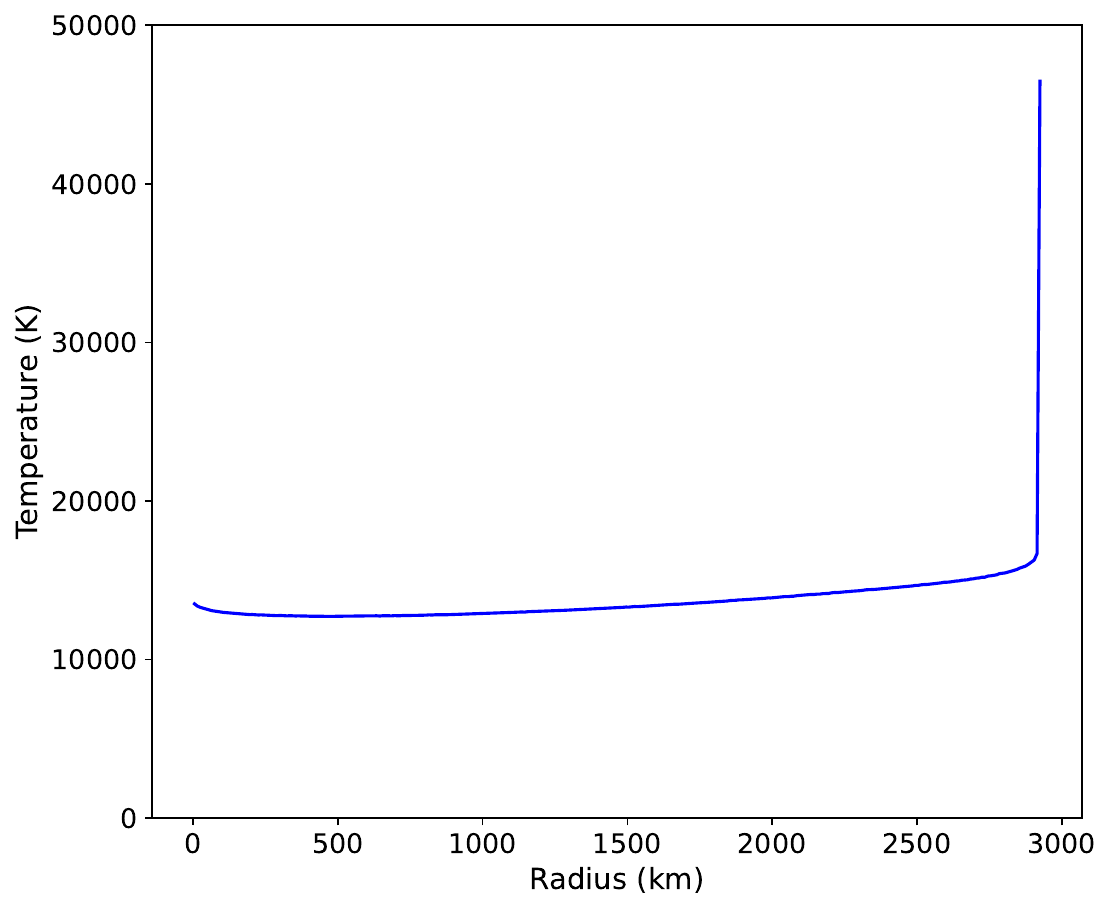}
    \end{tabular}
    \caption{
     Example of radial density, temperature, and ionization profiles for a nugget with mass $M_{{\rm nugget}} = 4 \cdot 10^{12}$g with mirror star core density $\rho_{{\rm core}} = 1 \rho_{{\rm core, \odot}}$ and heating rate parameter $\xi = 10^{-16}$, see \eref{e.Pheating2}. The total nugget luminosity is $6 \cdot 10^{-7} L_\odot$. At radii beyond the shown range, the nugget forms an unobservably faint and diffuse X-ray corona that is thermalized with the mirror star core, see discussion in \sref{s.stoppingcondition}.
    }
    \label{e.nuggetprofile}
\end{figure}

\subsection{Properties of Mirror Star Nuggets}
\label{s.properties}

\begin{figure*}
    \centering\newcommand{\tempsize}{2.8}
    \newcommand{\hpull}{\hspace*{-3mm}}    \newcommand{\hpulltwo}{\hspace*{-2mm}}
    \begin{tabular}{m{5mm}cccc}
    $\rho_{{\rm core}}$ & $T_{{\rm nugget}}(r = 0)$ (K) & $L/L_\odot$ & $r_{{\rm max} }$ (km) & $\log_{10} H_\alpha / NII$
    \\
    &
    \hpulltwo
    \includegraphics[width = 4 cm]{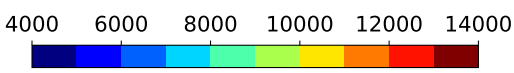}
    &
    \hpull
    \includegraphics[width = 4 cm]{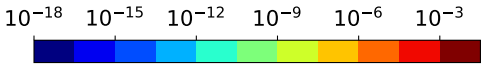}
    &
    \hpull
    \includegraphics[width = 4 cm]{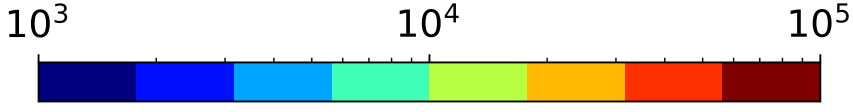}
    &
    \hpull
    \includegraphics[width = 4 cm]{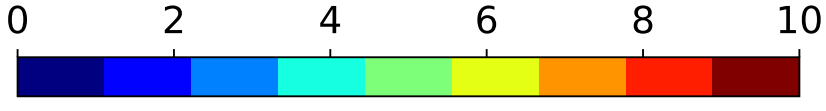}
    \\
    \rotatebox{90}{\phantom{aaa} $0.01 \rho_{{\rm core},\odot}$}
    &
    \hpulltwo
    \includegraphics[height=\tempsize cm]{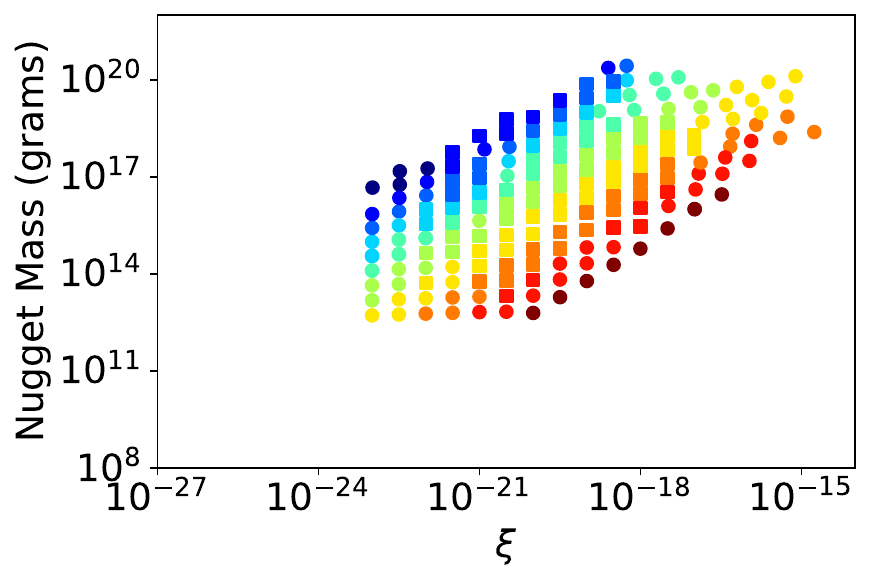}
    &
    \hpull
    \includegraphics[height=\tempsize cm]{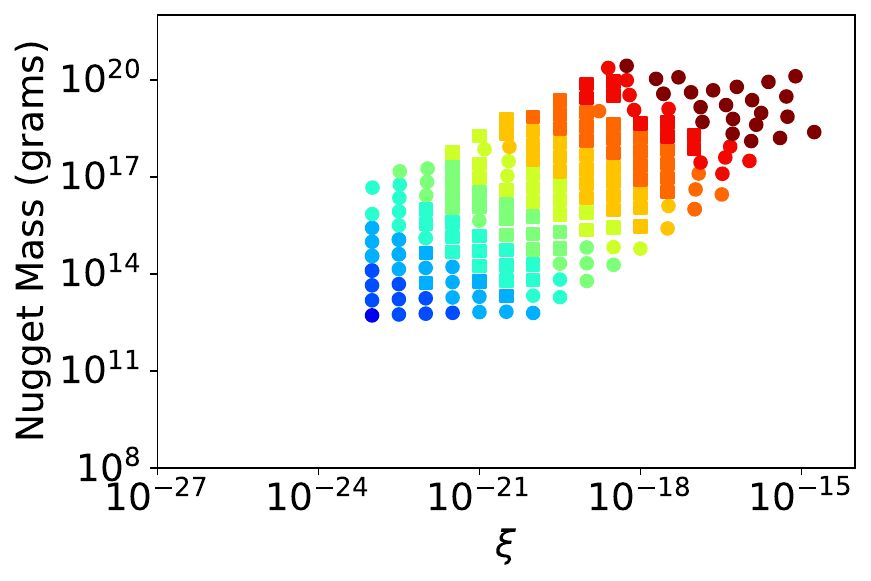}
    &
    \hpull
    \includegraphics[height=\tempsize cm]{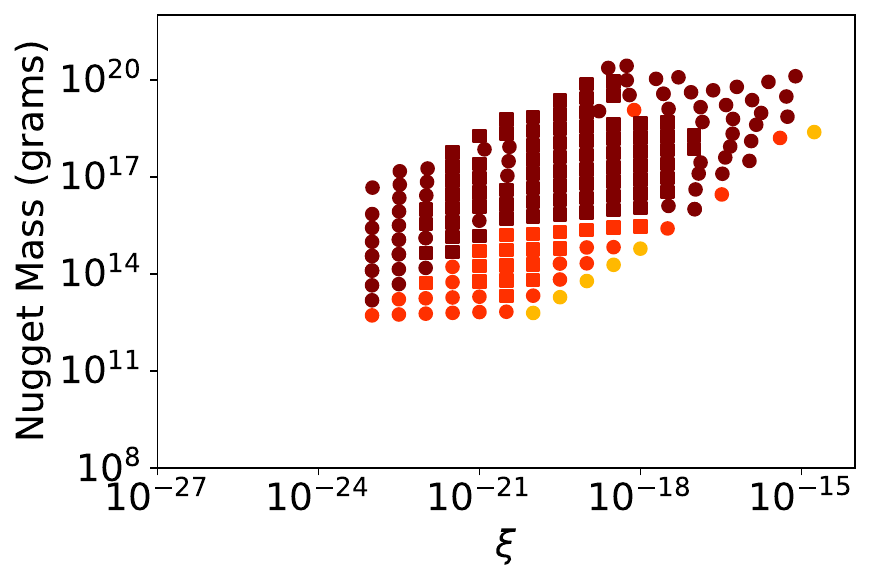}
    &
    \hpull
    \includegraphics[height=\tempsize cm]{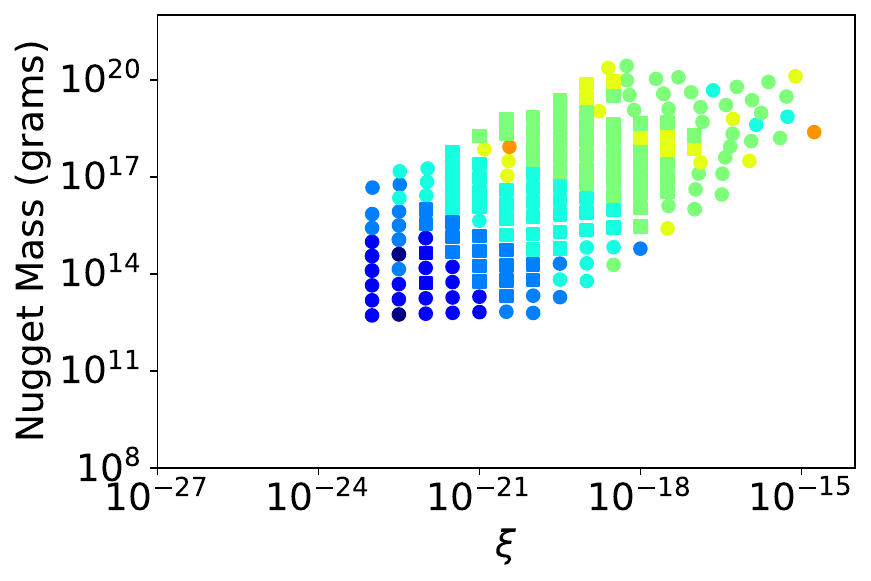}
    \\
    \rotatebox{90}{\phantom{aaaa} $0.1 \rho_{{\rm core},\odot}$}
    &
    \hpulltwo
    \includegraphics[height=\tempsize cm]{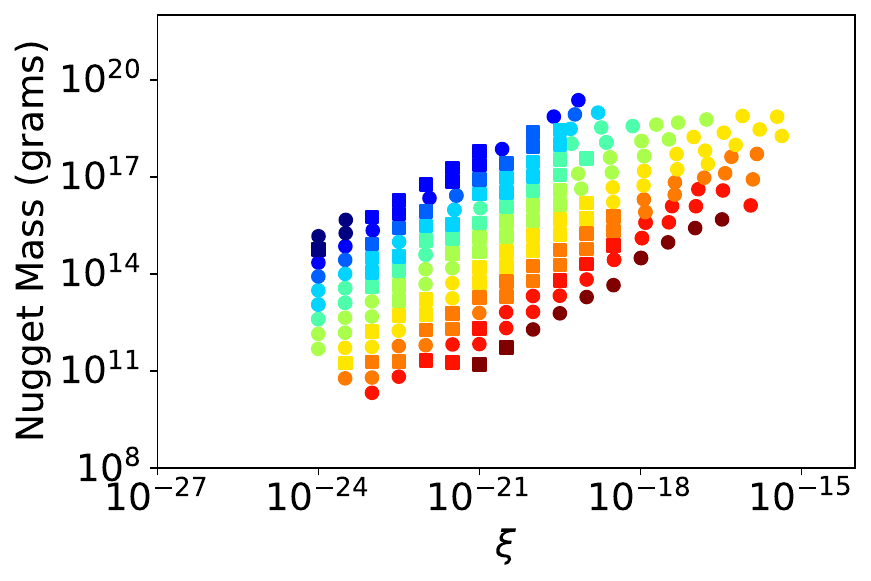}
    &
    \hpull
    \includegraphics[height=\tempsize cm]{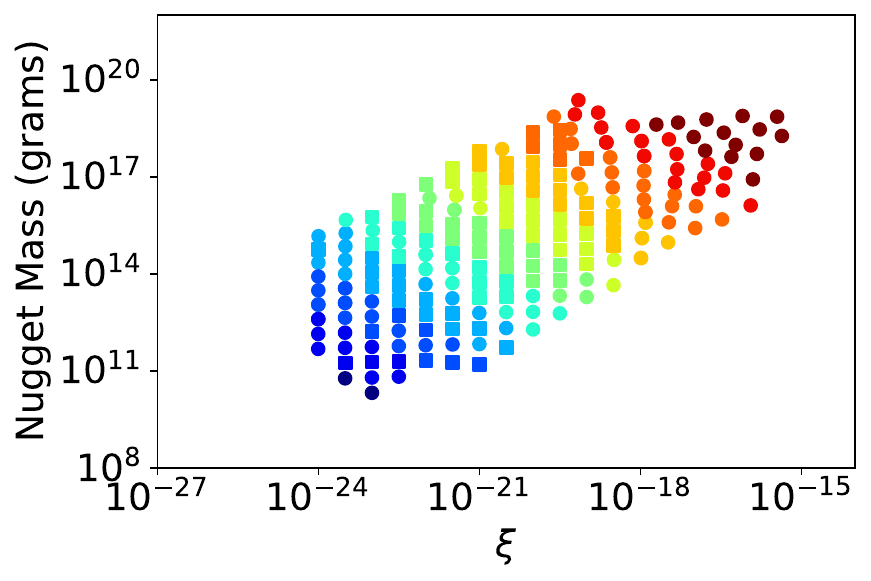}
    &
    \hpull
    \includegraphics[height=\tempsize cm]{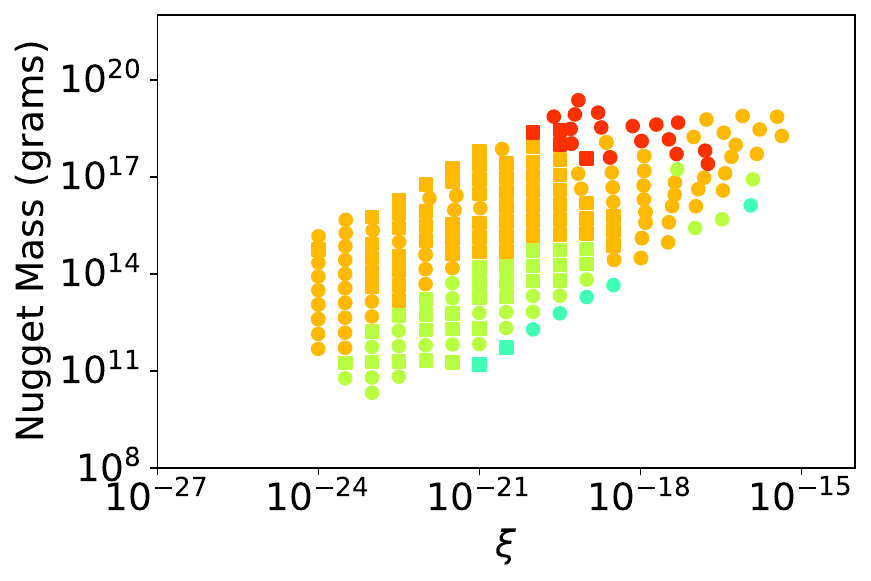}
    &
    \hpull
    \includegraphics[height=\tempsize cm]{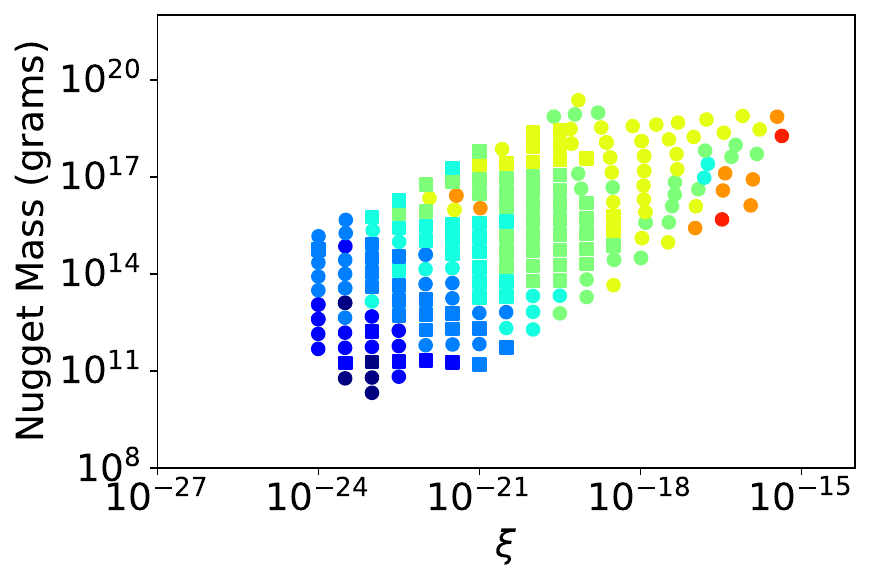}
    \\
    \rotatebox{90}{\phantom{aaaaaa} $ \rho_{{\rm core},\odot}$}
    &
    \hpulltwo
    \includegraphics[height=\tempsize cm]{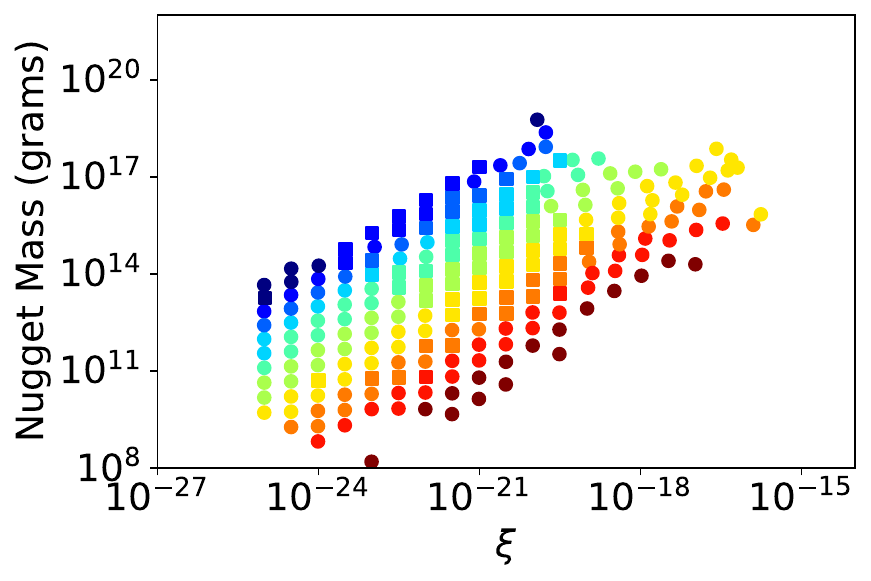}
    &
    \hpull
    \includegraphics[height=\tempsize cm]{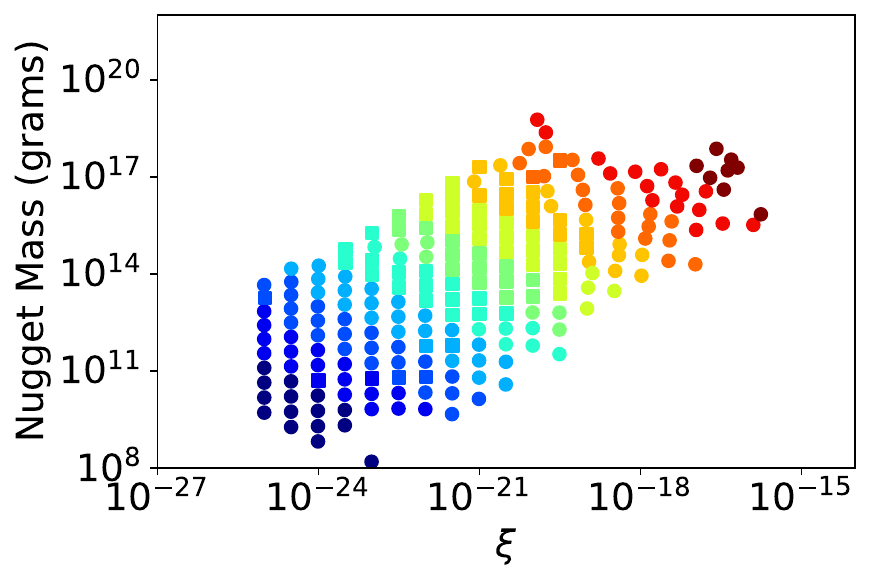}
    &
    \hpull
    \includegraphics[height=\tempsize cm]{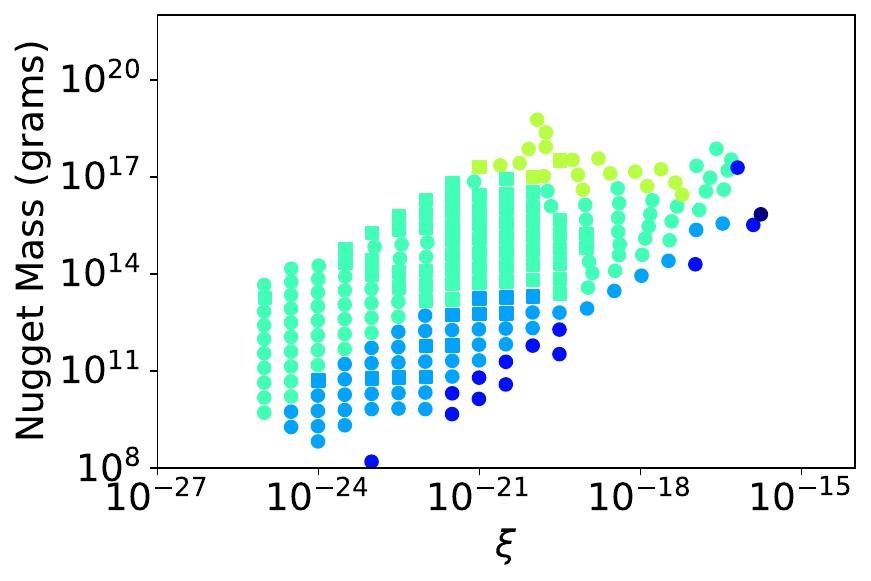}
    &
    \hpull
    \includegraphics[height=\tempsize cm]{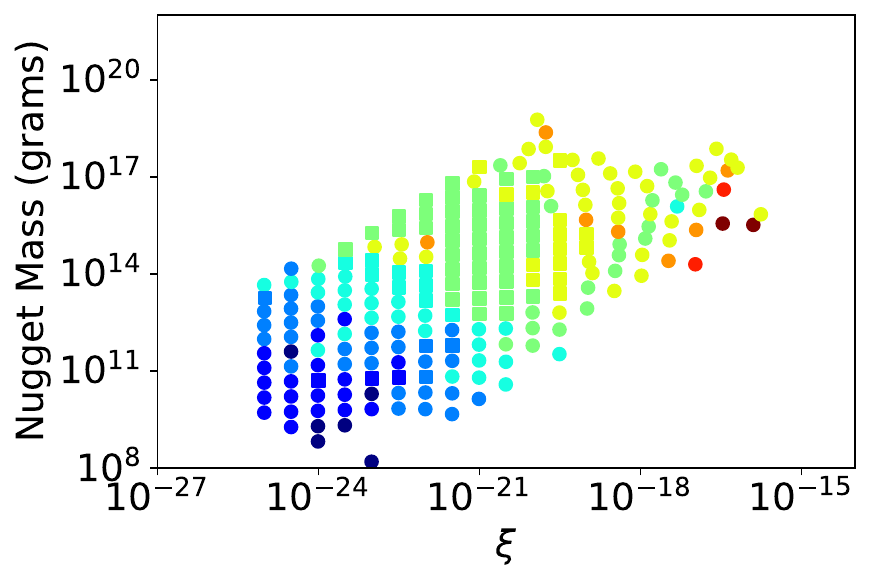}
    \\
    \rotatebox{90}{\phantom{aaaa} $10 \rho_{{\rm core},\odot}$}
    &
    \hpulltwo
    \includegraphics[height=\tempsize cm]{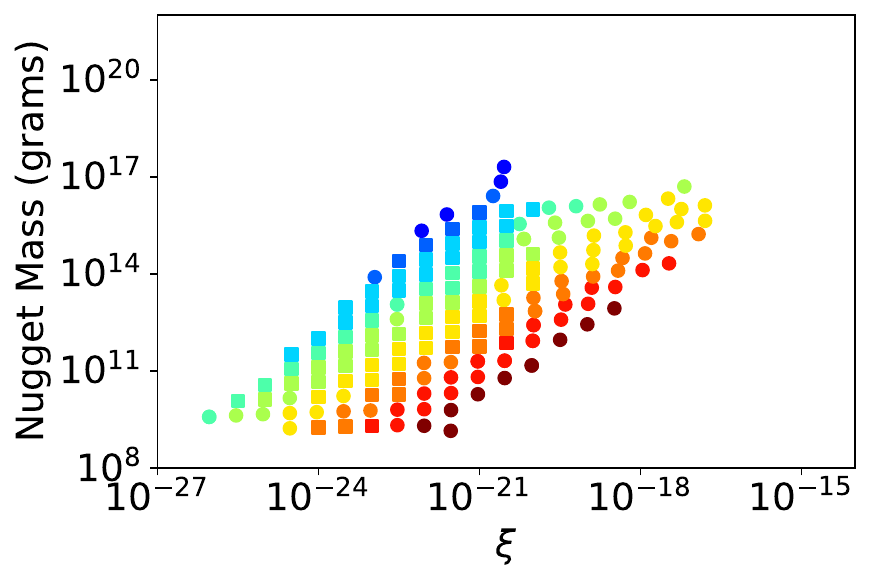}
    &
    \hpull
    \includegraphics[height=\tempsize cm]{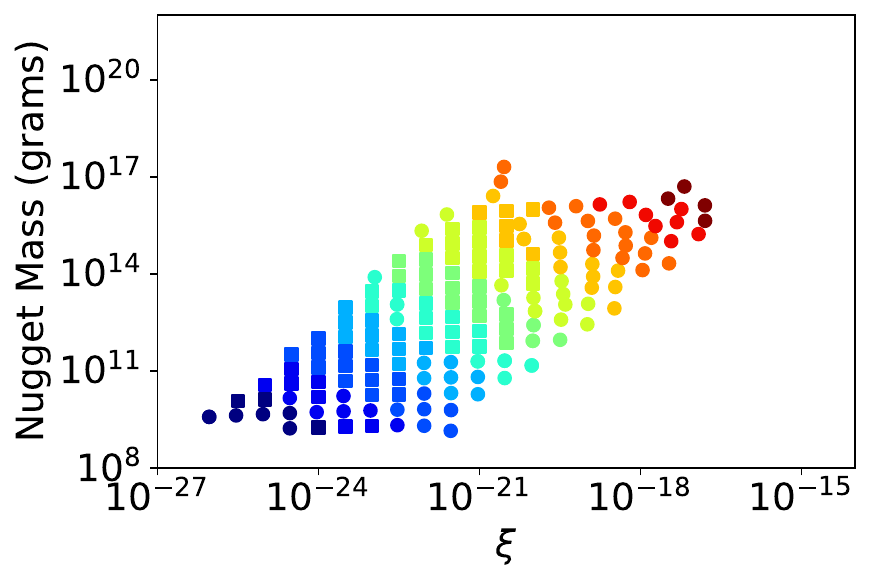}
    &
    \hpull
    \includegraphics[height=\tempsize cm]{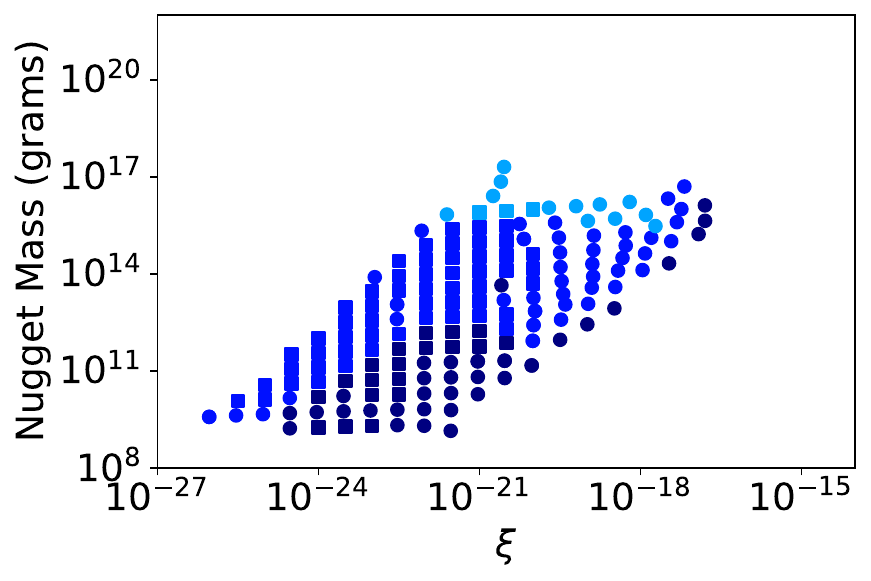}
    &
    \hpull
    \includegraphics[height=\tempsize cm]{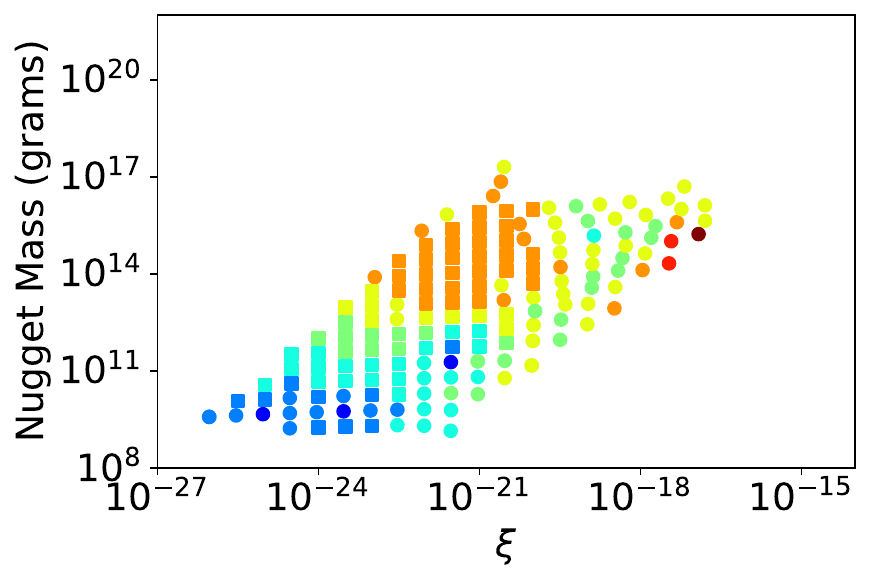}
    \\
    \rotatebox{90}{\phantom{aaa} $100 \rho_{{\rm core},\odot}$}
    &
    \hpulltwo
    \includegraphics[height=\tempsize cm]{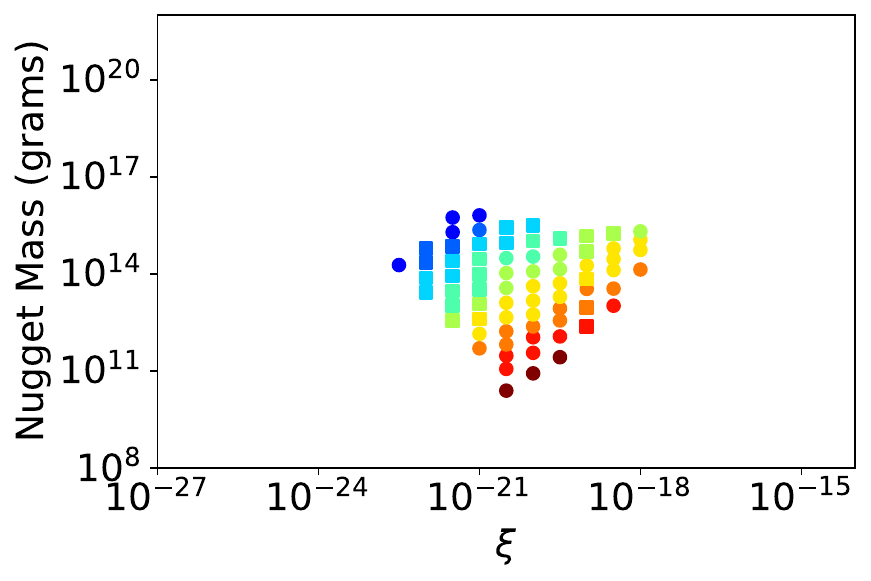}
    &
    \hpull
    \includegraphics[height=\tempsize cm]{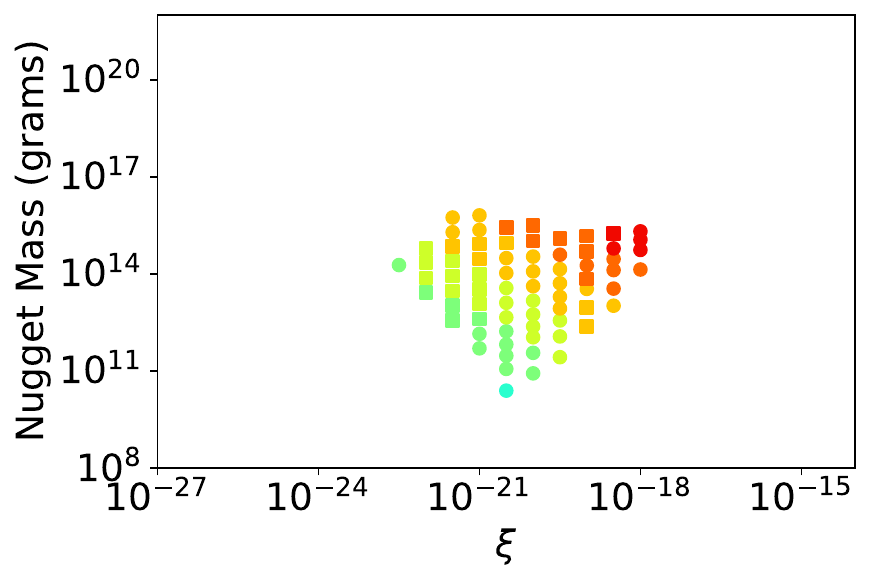}
    &
    \hpull
    \includegraphics[height=\tempsize cm]{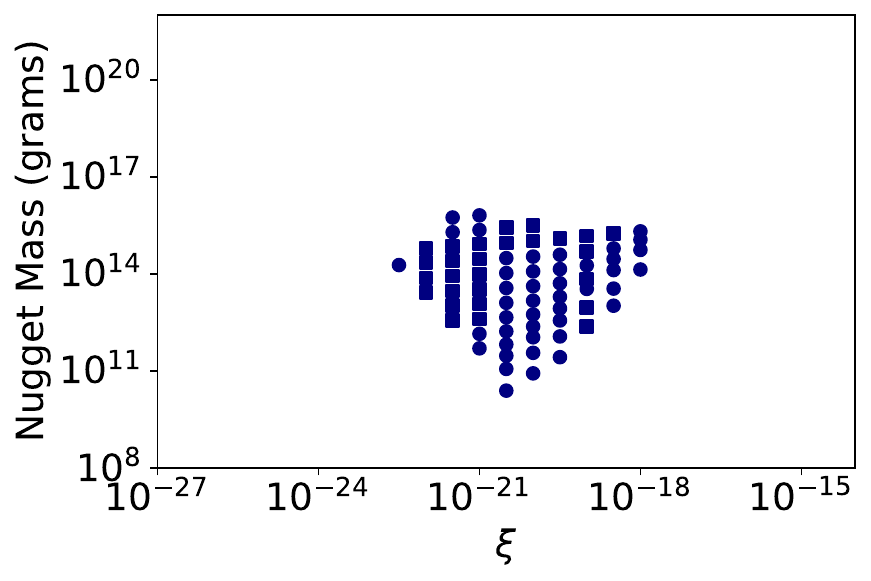}
    &
    \hpull
    \includegraphics[height=\tempsize cm]{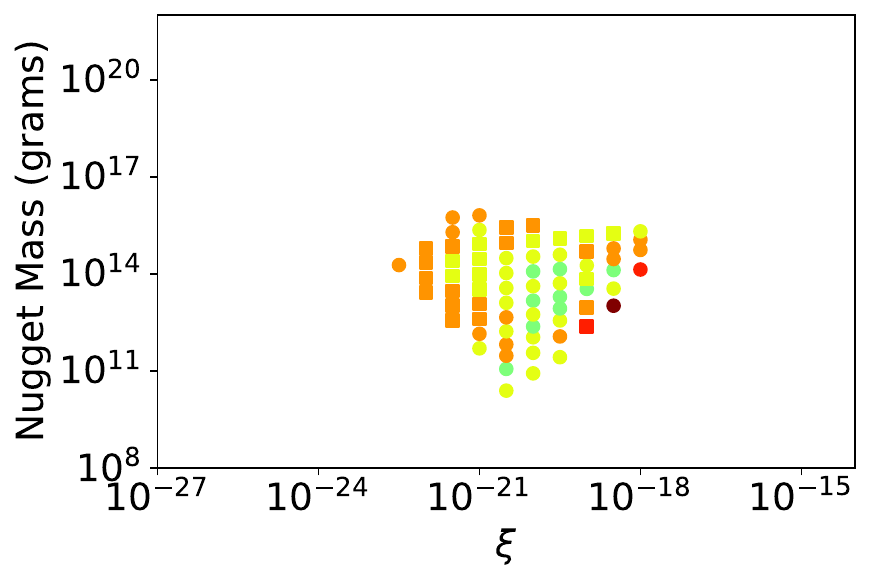}
    
    \end{tabular}
    \caption{
    Parameter space of optically thin mirror star nuggets in the plane of heating rate parameter $\xi$, see \eref{e.Pheating2}, and nugget mass  $M_{{\rm nugget}}$, for different $\rho_{{\rm core}} = (10^{-2},10^{-1},10^{0},10^1,10^2) \times \rho_{{\rm core}, \odot}$, with $\rho_{{\rm core}, \odot} \approx 160 g/cm^3$, in different rows from top to bottom. 
    Color indicates
    central nugget temperature (1st column),
    total nugget luminosity (2nd column), 
    effective nugget radius $r_{{\rm max}}$ (3rd column) and the emission line ratio H$\alpha$/N\,II (4th column).
    Circles are \Cloudy\ solutions, squares are obtained by interpolating the shown physical nugget property across the gap where \Cloudy\ cannot complete the integration.  
    }
    \label{f.paramspace}
\end{figure*}

We show a representative nugget solution produced by \texttt{Cloudy} in \fref{e.nuggetprofile}. 
The scale height of the density column is $\sim 10^3$\,km 
as expected from \eref{e.hnugget}. Temperature is lowest near the center, where the transparent nugget can cool most efficiently due to the higher central density, but increases only very modestly until close to $r_{{\rm max}}$, where it jumps sharply towards the integration limit $T_{{\rm max}} = 8 \times 10^4$\,K. The unobservably faint and diffuse X-ray corona exists beyond $r_{{\rm max}}$.

We run \Cloudy\ for mirror star core densities $\rho_{{\rm core}} = (10^{-2},10^{-1},10^{0},10^1,10^2) \times \rho_{{\rm core}, \odot}$.
For each $\rho_{{\rm core}}$, nugget solutions are found in a grid of central hydrogen density $n_H$ (determining $M_{{\rm nugget}}$) and heating rate parameter $\xi$, and our range of parameters covers the entire observable range where \texttt{Cloudy} is able to find stable equilibrium nugget solutions.
This parameter space is shown in \fref{f.paramspace}.

Away from our scan points in \fref{f.paramspace}, nuggets generated by \texttt{Cloudy} fail to pass various consistency checks, either \Cloudy's  or those we impose in our analysis. 
Specifically, the white region in the upper left corresponds to optically thick nuggets where \Cloudy\ cannot exactly solve for full radiative balance and hydrostatic equilibrium. In the upper right, 
\Cloudy\ generates seemingly stable solutions, but we find that they do not obey global energy balance, i.e., the total luminosity significantly deviates from the total heating rate. This is likely due to the initial central density being too high for \Cloudy\ to self consistently solve. Although there is no strict upper density limit imposed by \Cloudy, its prescriptions for self consistency break down in the optically thick limit. This explains why \Cloudy\ can generate a full solution while failing our additional consistency checks.
We impose the requirement that total luminosity and total heating rate agree within a factor of 2. For nuggets where this ratio is different from 1 but within a factor of 2, we shift the nugget's assumed $\xi$-value in the parameter space to agree with its luminosity. This results in some  deviations from a regular grid near the top right corners of the plots in \fref{f.paramspace}.

In the lower right, the density is so low and the heating rate so high that nugget temperatures violate the non-relativistic assumptions of \texttt{\texttt{Cloudy}}.  On the other hand, nuggets in the lower left are unobservably faint, which informs the limits of our scan range. For mirror star core densities of $0.1 \rho_\mathrm{core, \odot}$ or higher, \texttt{Cloudy} cannot find stable solutions for the lowest heating rates and nugget masses we explore, likely due to the steep dependence of the cooling rate at very low temperatures in \fref{f.cooling}.
Finally, we also remove any \texttt{Cloudy} solutions that only have three or fewer solved zones. These nuggets lie along the diagonal boundary of the lower right corner of our parameter space, and reach \Cloudy's temperature limit of 
$10^9$\,K within a few zones. This is due to the balance between the nugget heating $\xi$, and the ability of the nugget to efficiently cool. 
These regions correspond to extremely low-mass nuggets that are never observable in practice, and would represent only an extremely transient early stage in nugget evolution as the mirror star accumulates more SM matter. 

For the remaining solutions, we show how central nugget temperature, total nugget luminosity, nugget radius before transitioning to the corona at $r_{max}$, and the H$\alpha$/N\,II  emission line ratio, vary with $(\xi, M_{{\rm nugget}})$. These quantities vary essentially as expected across this parameter space. Nugget temperature will increase as $\xi$ increases, but will decrease as mass increases at a fixed $\xi$ due to the increased cooling efficiency with increased nugget density.
Nugget luminosity increases as $(\xi, M_{nugget})$ increase, and the trend is uniform across different values of $\rho_{core}$. Larger values of $\rho_{core}$ means the nugget is situated in a stronger gravitational potential generate, and therefore have a smaller radii. The H$\alpha$/N\,II  emission line ratio becomes larger with increasing nugget luminosity, as H$\alpha$ emission is the dominant cooling channel.

\Cloudy\ raises an error if the Compton optical depth $\int_0^{r_{\rm max}} \sigma_{\rm es} n_e dr$ becomes large.  
We additionally compare the nugget's luminosity to that of an optically thick blackbody of similar size and temperature. Valid solutions should satisfy $L_{\rm nugget}/(4\pi R_{\rm nugget}^2\sigma T_{\rm nugget}^4) \ll 1$, and we have confirmed that this is indeed the case for nuggets that pass the various consistency checks described above.

Here we must point out that there is another region where \texttt{Cloudy} does not produce stable nugget solutions, the ``gap'' at moderate heating rates and masses, identified with square markers in \fref{f.paramspace}. 
This region is surrounded by regions where \texttt{Cloudy} returns stable solutions, and after thorough investigation we were unable to identify any physical reason why the nuggets in this gap would not be optically thin and solvable by \texttt{\texttt{Cloudy}}.
We suspect that an unknown numerical issue causes the code to fail in these regions, and we deal with this issue by interpolating the various physical nugget properties (including $M_{nugget}$, which is derived from the central density initial condition) of interest across the gap using Delaunay triangulation. All physical characteristics vary very smoothly across the nugget parameter space, seemingly with powers of linear combinations of $M_{{\rm nugget}}$ and $\xi$, and the interpolation output does not significantly depend on the exact method of interpolation employed.\footnote{Some of the emmission line ratio predictions for heavy nuggets have slightly non-trivial variation across the parameter space, and therefore interpolation could introduce some quantitative uncertainty, but the differences between the line ratios of nuggets and planetary nebulae are so drastic that this will not affect any of our conclusions.}
While first-principles \texttt{Cloudy} solutions would obviously be preferable, we find this to be a satisfactory solution, and clearly mark those parameter points obtained by interpolation in subsequent plots.

Generally, higher mirror star core densities yield fewer observable optically thin nuggets: with total nugget luminosity determined by its mass, higher core densities means nuggets of the same mass are denser and therefore more likely to be optically thick, placing them
beyond the scope of our current analysis. For $\rho_{{\rm core}} = 100 \rho_{{\rm core}, \odot}$ we find very few optically thin nuggets with magnitude $< 30$, meaning considering higher core densities would not yield additional observable optically thin nuggets. Of course, we could consider core densities below our minimum of $10^{-2} \rho_{{\rm core},\odot}$, but this would merely continue the trends in nugget observables that will be clearly identified in the next section.

\begin{figure}
    \centering
    \begin{tabular}{l}
    \includegraphics[width = 5.5 cm]{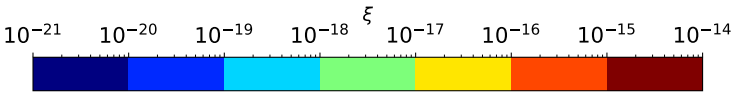} 
    \\
    \includegraphics[width = 5.5 cm]{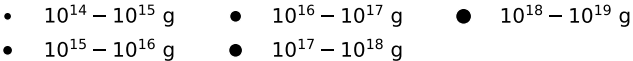}
    \\
    \rotatebox{90}{\phantom{aaaa} $\rho_\mathrm{core} = 0.01 \rho_{{\rm core},\odot}$}
    \includegraphics[trim=0 5mm 0 18mm,clip, height=3.7cm]{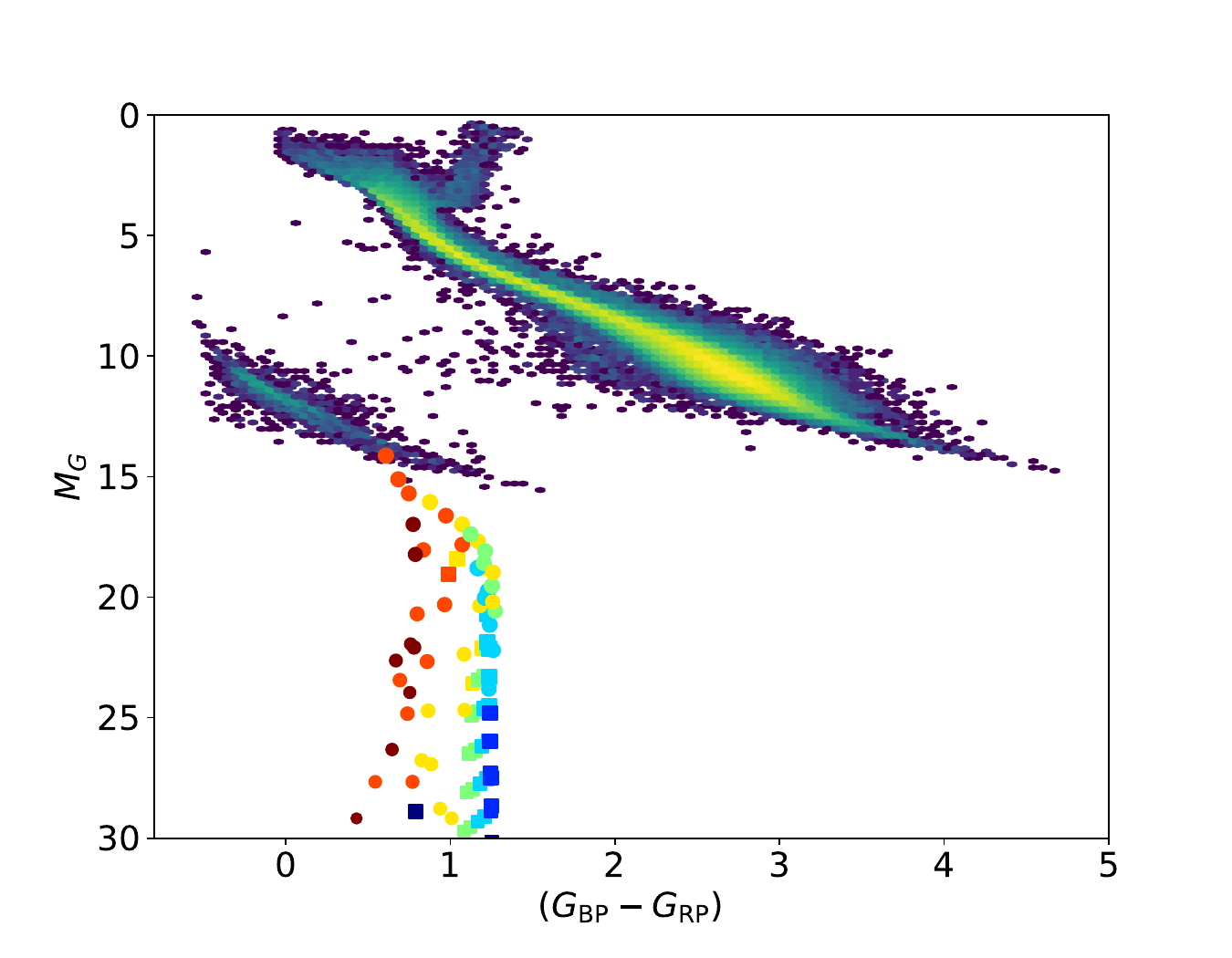} 
    \\
    \rotatebox{90}{\phantom{aaaaa} $\rho_\mathrm{core} = 0.1 \rho_{{\rm core},\odot}$}
    \includegraphics[trim=0 5mm 0 18mm,clip, height=3.7cm]{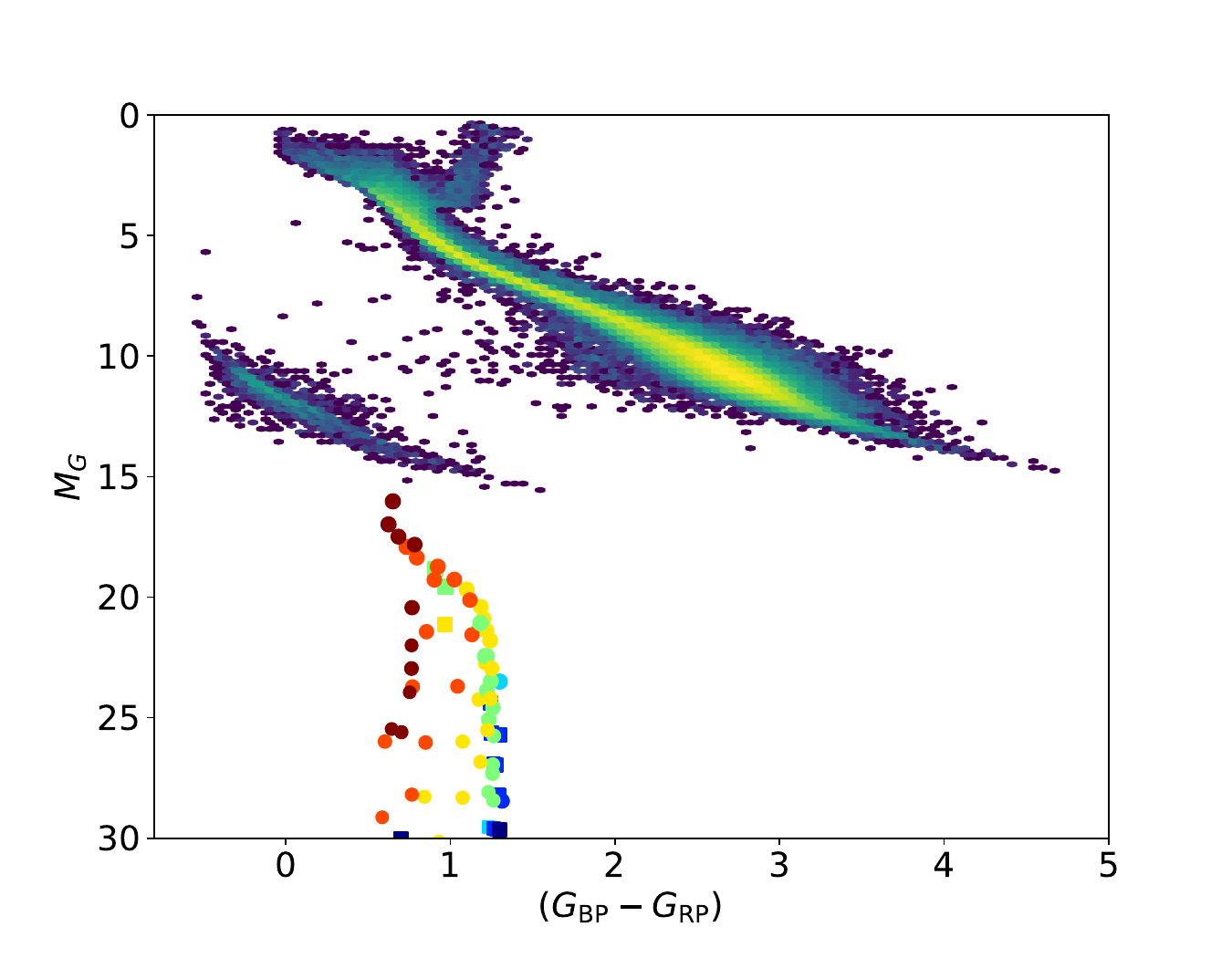}
    \\
    \rotatebox{90}{\phantom{aaaaaa} $\rho_\mathrm{core} = 1 \rho_{{\rm core},\odot}$}
    \includegraphics[trim=0 5mm 0 18mm,clip, height=3.7cm]{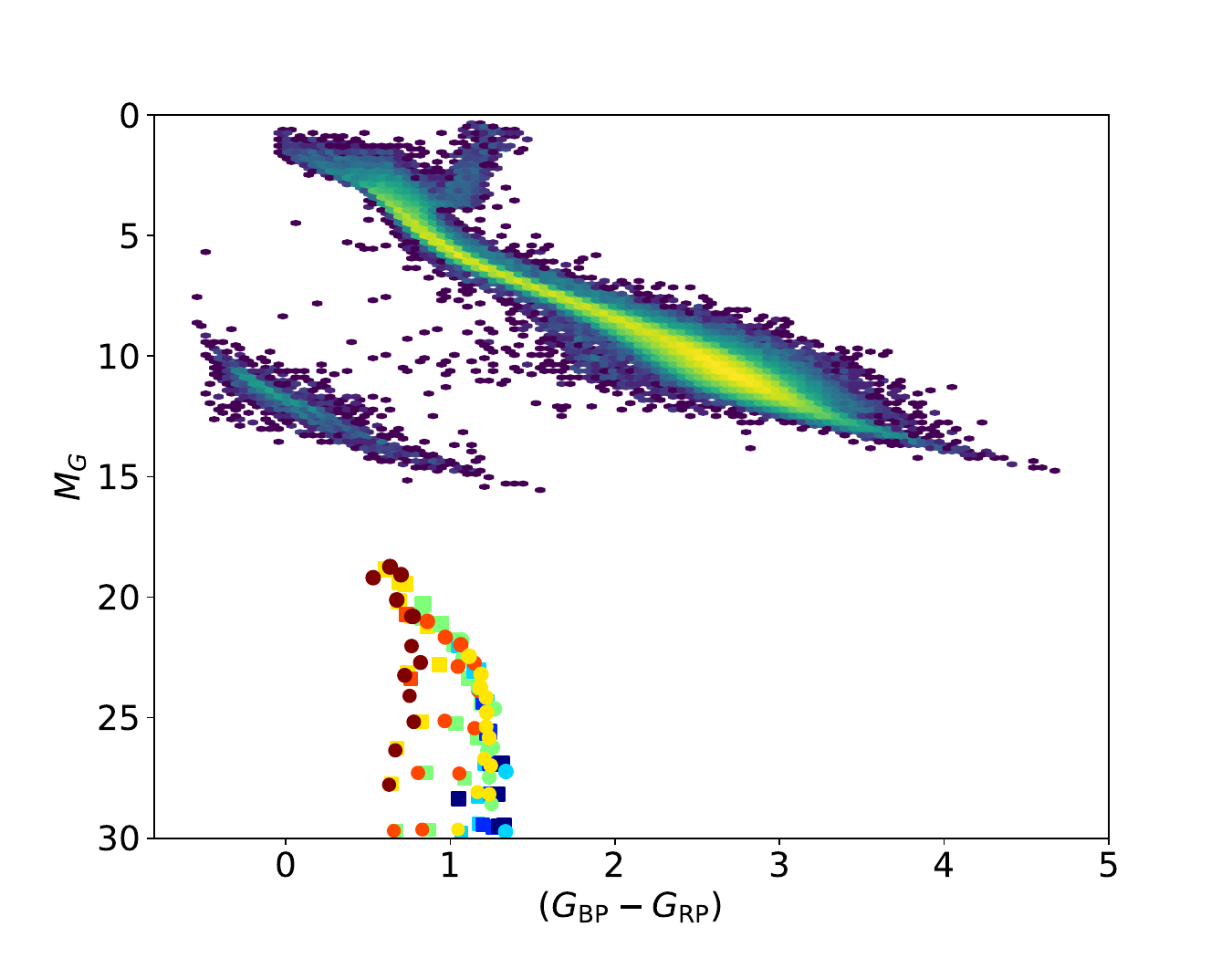}
    \\
    \rotatebox{90}{\phantom{aaaaa} $\rho_\mathrm{core} = 10 \rho_{{\rm core},\odot}$}
    \includegraphics[trim=0 5mm 0 18mm,clip, height=3.7cm]{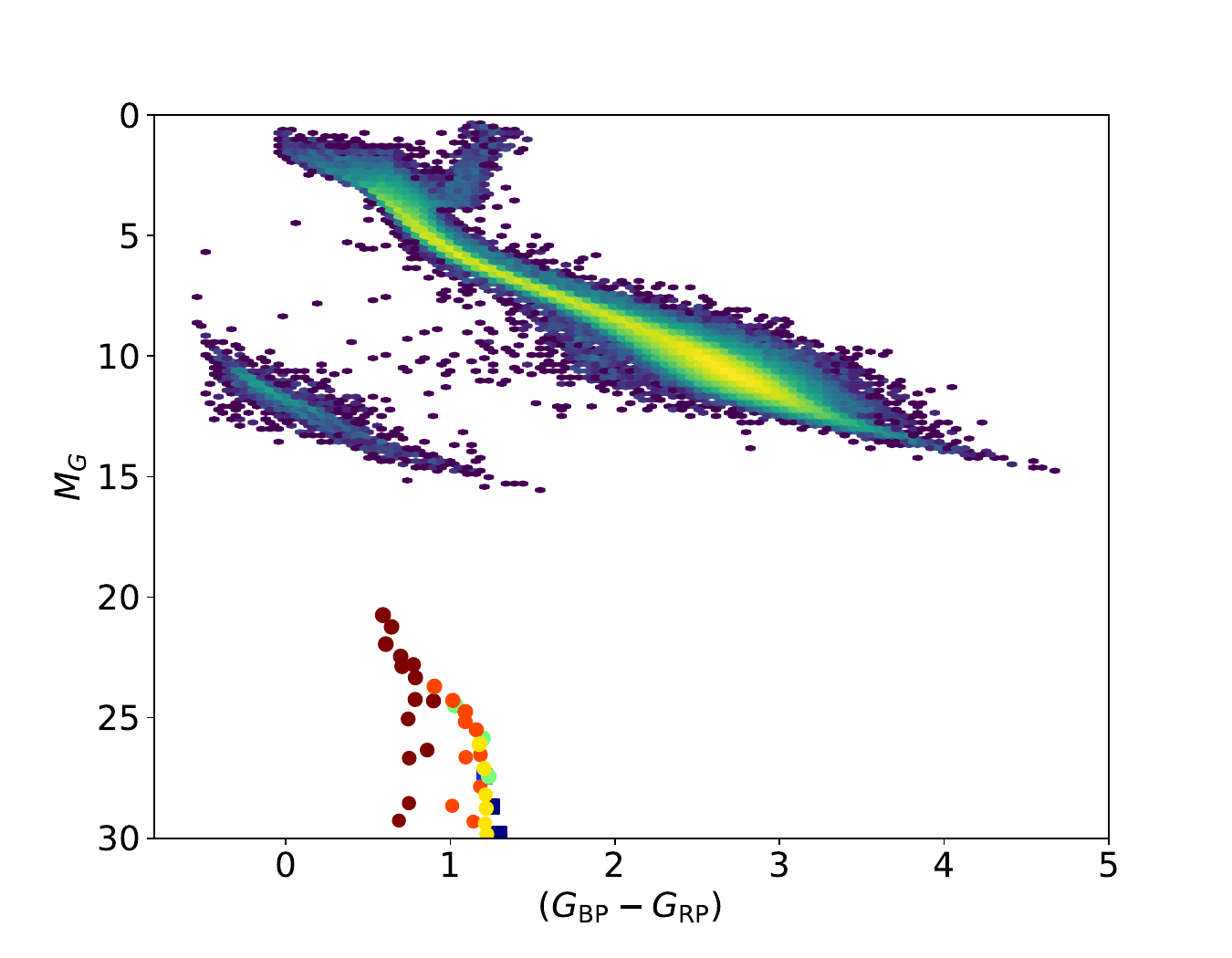}
    \\
    \rotatebox{90}{\phantom{aaaa} $\rho_\mathrm{core} = 100 \rho_{{\rm core},\odot}$}
    \includegraphics[trim=0 5mm 0 18mm,clip, height=3.7cm]{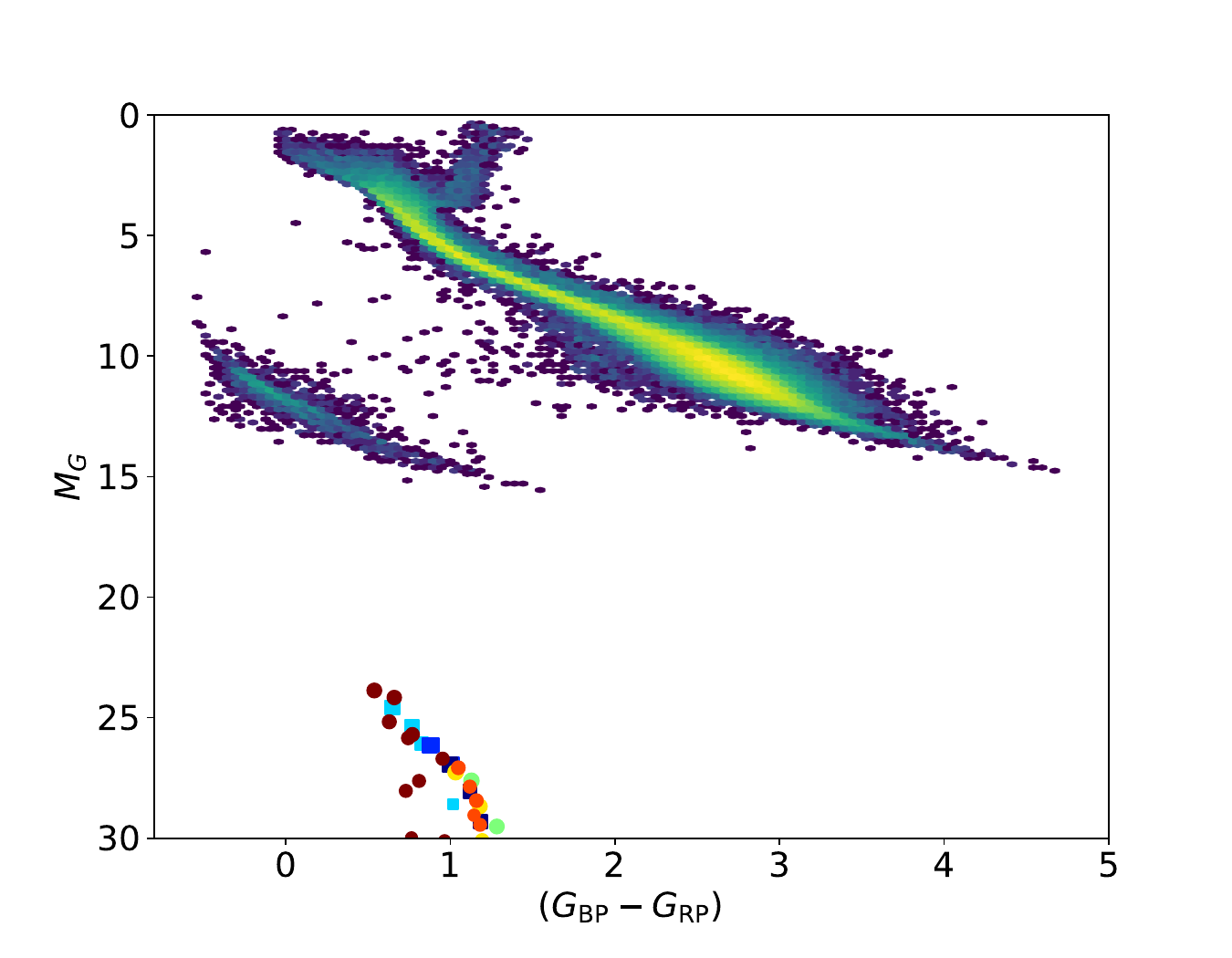}
    \end{tabular}
    \caption{
    Hertzsprung-Russell diagram of sources from the \Gaia DR3 catalogue and optically thin mirror star nuggets for different mirror star core densities. For the nuggets, marker color (size) indicates heating rate $\xi$ ($M_{{\rm nugget}}$). Circles are Cloudy solutions to nuggets, whereas squares are interpolated.
    }
    \label{f.HR}
\end{figure}

\section{Electromagnetic Mirror Star Signatures}
\label{s.results}
We now describe the electromagnetic signatures of optically thin nuggets in Mirror Stars from near-ultraviolet to near-infrared
wavelengths. 
This will establish a clear mirror star signal region in the HR diagram, but importantly also establish purely spectral methods of distinguishing mirror stars from other astrophysical objects like white dwarfs and planetary nebulae. 

\subsection{Distribution of Mirror Stars in HR Diagram}
\label{s.HR}

A Hertzsprung–Russell (HR) diagram comprised of stars with parallex of 1 to 80 mas and error less than 5\%  selected from the \Gaia DR3 is shown in \fref{f.HR}. We show nuggets brighter than $G = 30$, with dimmer nuggets deemed unobservable. The \Gaia passband functions \citep{Jordi10GaiaPassbands} were used to obtain color and absolute magnitude values for our optically thin mirror star nuggets and place them in the HR diagram, with marker color (size) indicating heating rate $\xi$  ($M_{{\rm nugget}}$). 
As expected, 
optically thin nuggets have lower maximum luminosity for higher mirror star core densities, since they transition to being optically thick at lower masses/luminosities. 
These optically thick nuggets will be studied in a future investigation.

Remarkably, there is an extremely well-defined ``mirror star signal region''. Regardless of $\rho_{{\rm core}}$, 
the optically thin nuggets populate a narrow band in colour space,
\begin{equation}
    G_{BP} - G_{RP} \in (0.4, 1.4) \ .
\end{equation}
In absolute magnitude, the nuggets can range from arbitrarily low luminosity up to the luminosities of white dwarfs.  Note that these luminous objects must be at least moderately optically thick.
This is notably different from the signal region assumed in the  \Gaia search demonstration~\cite{Howe:2021neq}, but this is expected due to the toy-emission model employed in that early analysis. The results of our work would now enable a realistic mirror star search in \Gaia data to be conducted. 

The position of an optically thin nugget in the HR diagram does not directly reveal its physical properties, though nugget brightness scales with total heating rate and nugget mass and lower heating rates tend to be clustered towards the redder end of the signal region.

\subsection{Detailed 
Nugget Emission Spectra}
\label{s.spectra}

There are two important reasons to study the detailed spectra of optically thin mirror star nuggets. First, the mirror star region of the HR diagram overlaps  that of white dwarfs, making distinguishing the two an observational priority. Second, it would be desirable to identify purely spectral methods of identifying mirror stars and distinguishing them from standard astrophysical point sources, without requiring  a parallax measurement. This would enable mirror star searches using spectroscopic surveys and other instruments that may have much greater sensitivity than \texttt{Gaia}.

\begin{figure}[t!]
    \centering
    \hspace*{-12mm}
    \begin{tabular}{c}
    \includegraphics[width=0.5\textwidth]{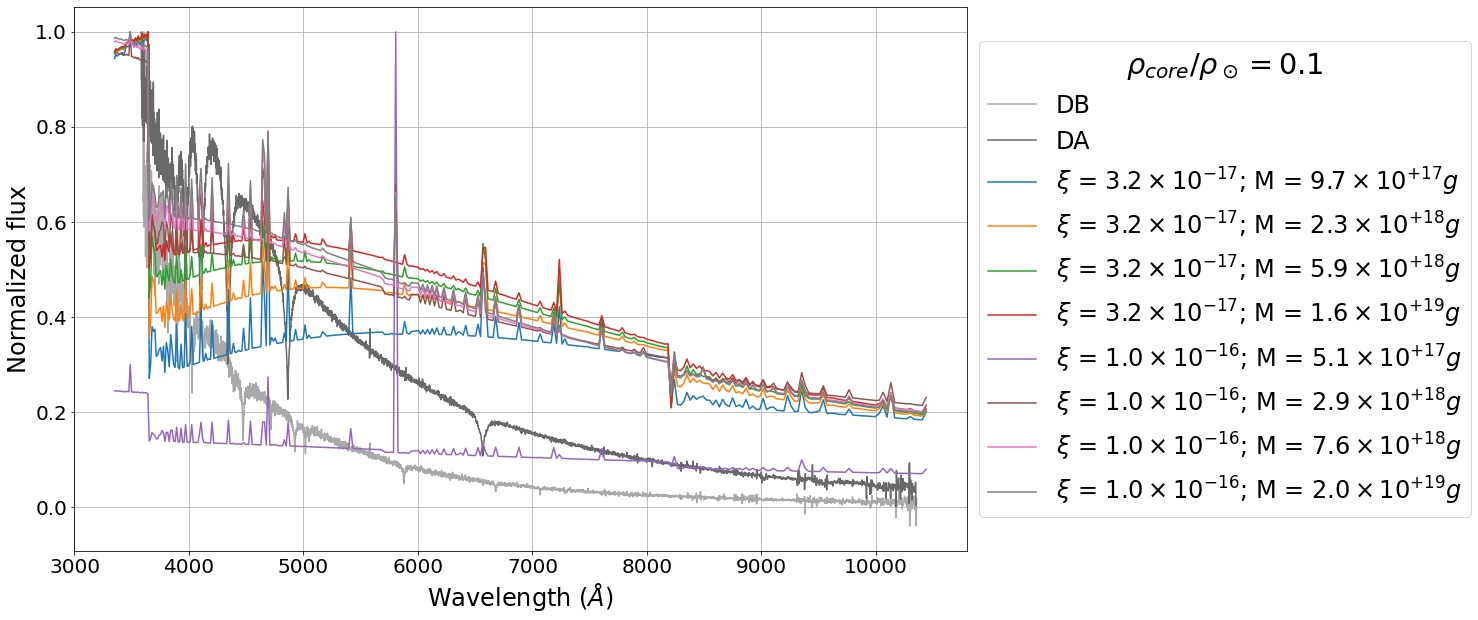}
    \\
    \includegraphics[width=0.5\textwidth]{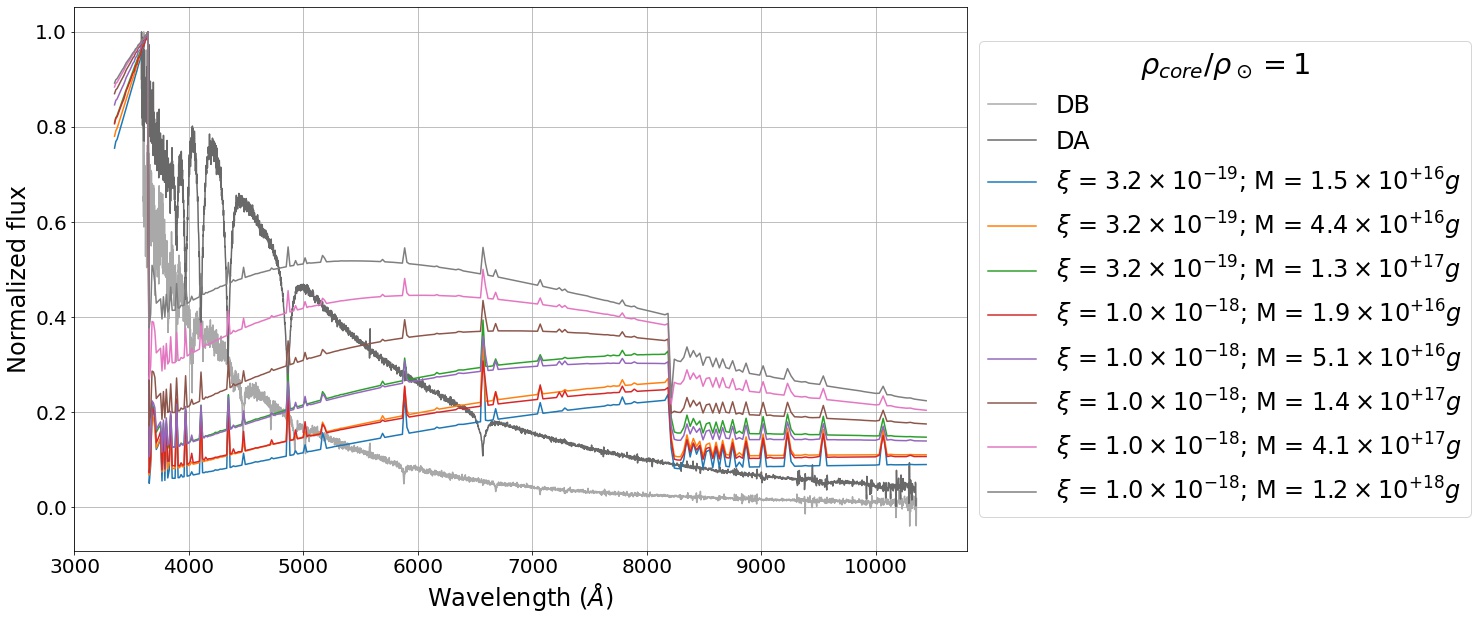}
    \\
    \includegraphics[width=0.5\textwidth]{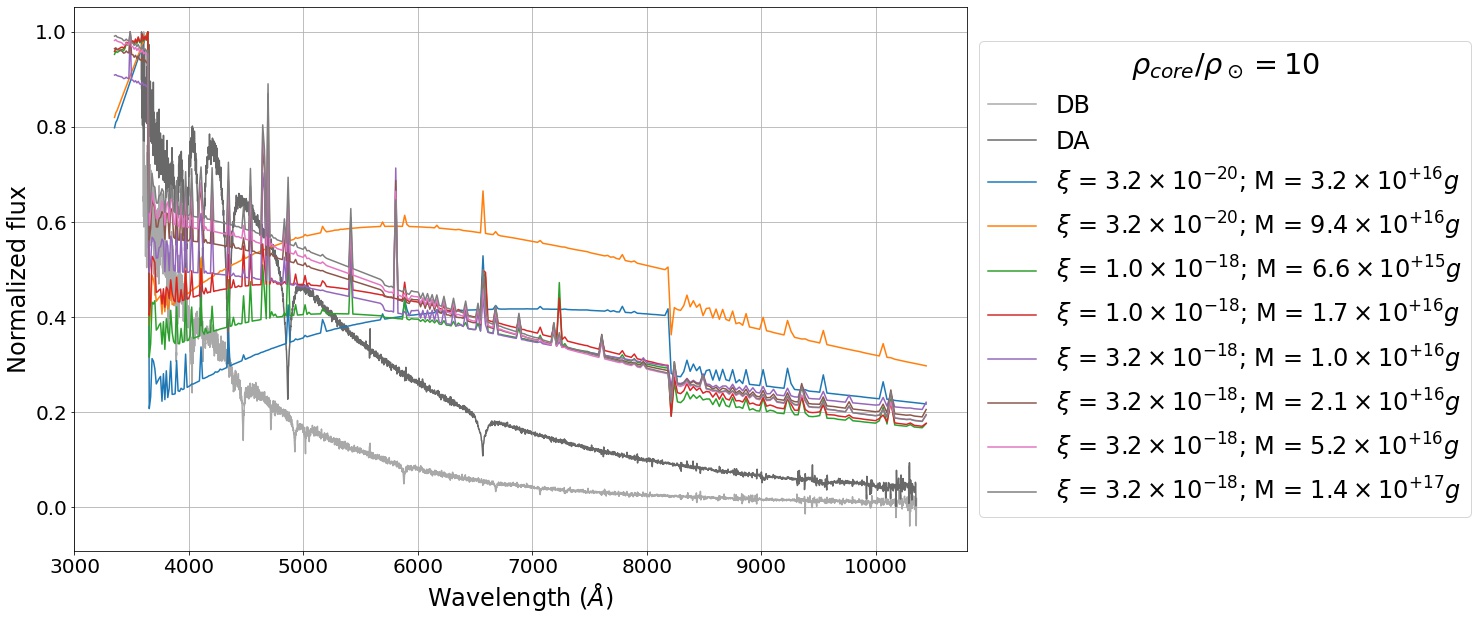}
    \end{tabular}
    \caption{
    Comparison of several representative white dwarf spectra from the SDSS (BOSS) public data set with emission spectra from observable optically thin nuggets captured by mirror stars. The difference between the black-body-like white dwarf spectrum shape and the continuum of the nuggets is apparent, as are the prominent Balmer and Paschen jumps at 3645\,\AA\ and 8206\,\AA, respectively. 
    }
    \label{f.spectra}
\end{figure}

\begin{figure}
    \centering
    \newcommand{\templengthbla}{2.8}
    
    \hspace*{-8mm}
    \begin{tabular}{cc}
    \hspace*{-6mm}
    \includegraphics[width = 5.2 cm]{figures/Colorbar_HR.png}
    &
    \hspace*{-6mm}
    \includegraphics[width = 4 cm]{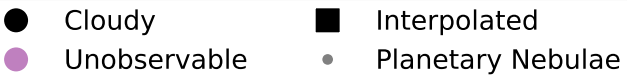}
    \\
    \rotatebox{90}{\phantom{a} $\rho_\mathrm{core} = 0.01 \rho_{{\rm core},\odot}$}
    \includegraphics[height=\templengthbla cm]{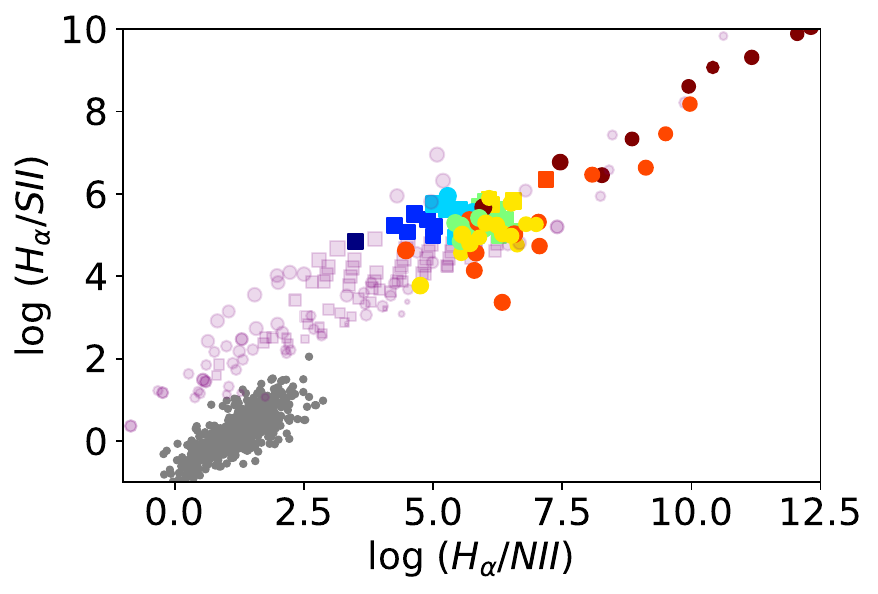}
    &
    \hspace*{-6mm}
    \includegraphics[height=\templengthbla cm]{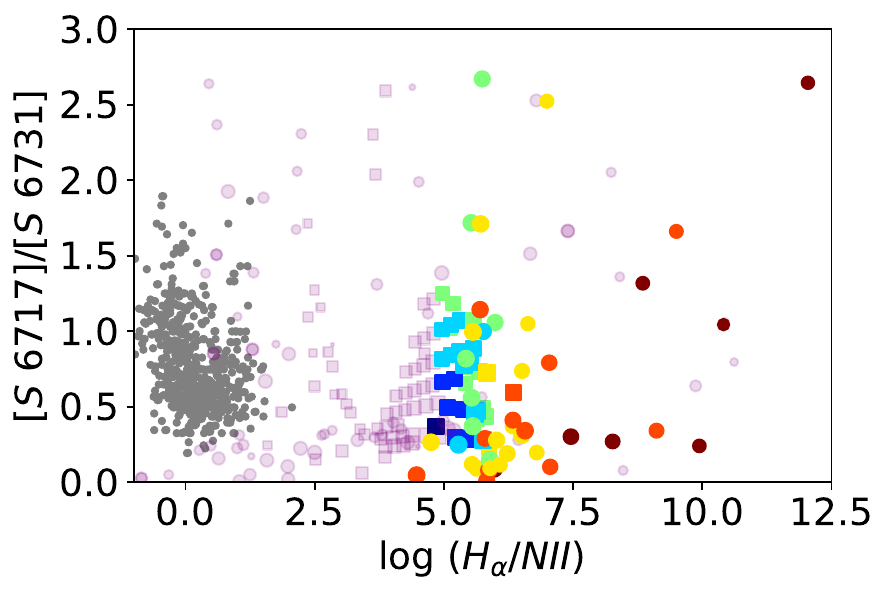}

    \\
    \rotatebox{90}{\phantom{aa} $\rho_\mathrm{core} = 0.1 \rho_{{\rm core},\odot}$}
    \includegraphics[height=\templengthbla cm]{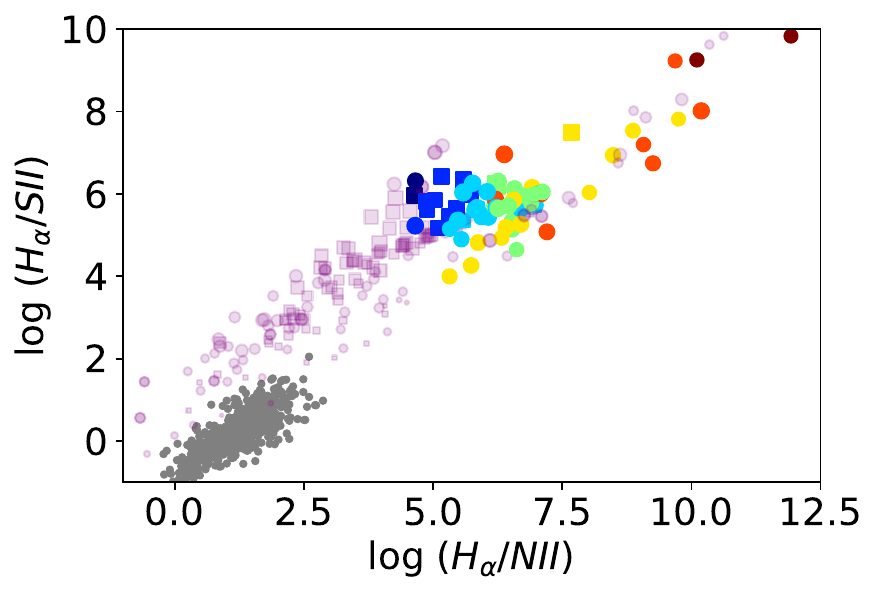}
    &
    \hspace*{-6mm}
    \includegraphics[height=\templengthbla cm]{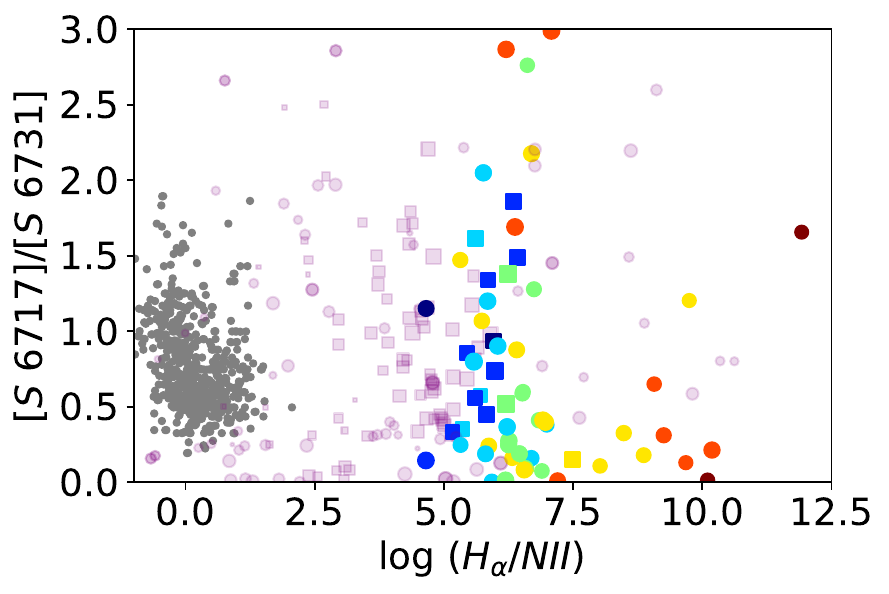}
    
    \\
    \rotatebox{90}{\phantom{aaa} $\rho_\mathrm{core} = 1 \rho_{{\rm core},\odot}$}
    \includegraphics[height=\templengthbla cm]{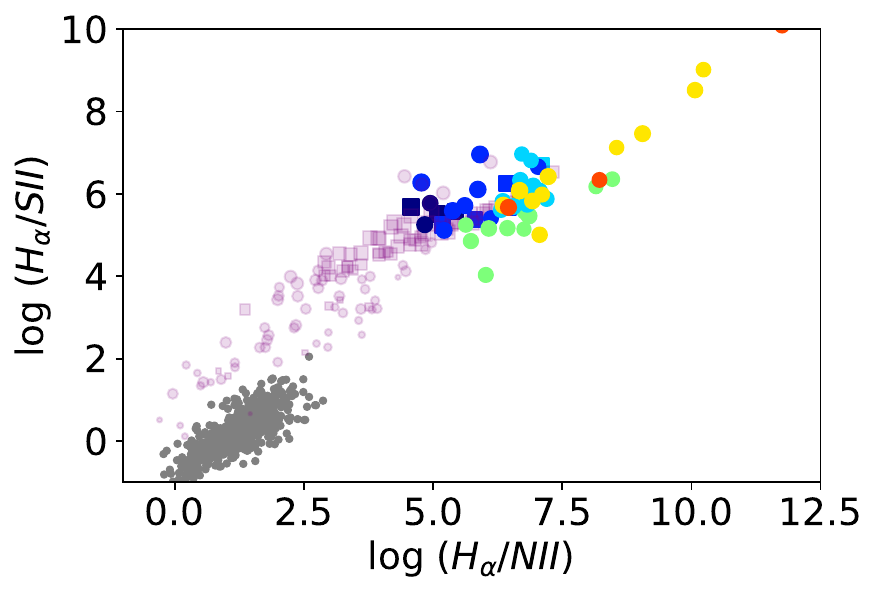}
    &
    \hspace*{-6mm}
    \includegraphics[height=\templengthbla cm]{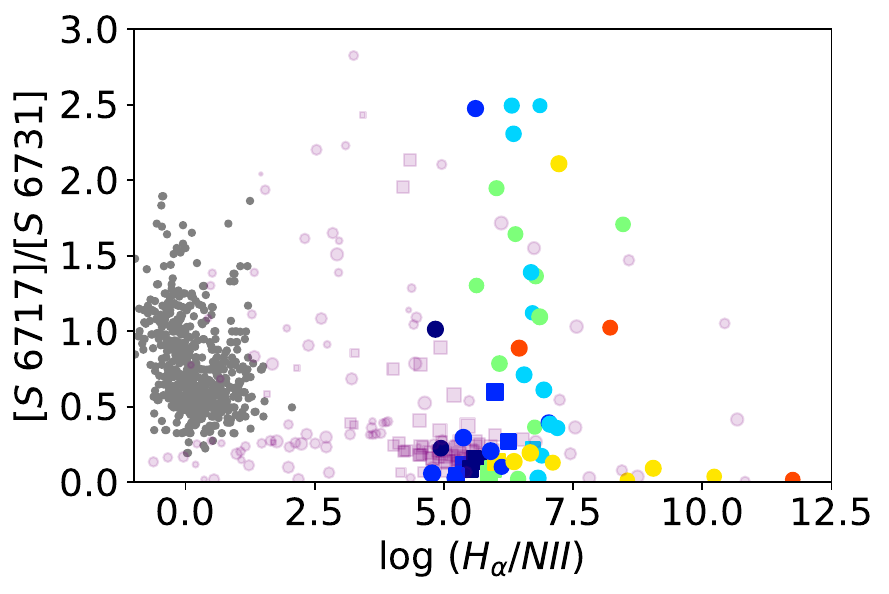}

    \\
    \rotatebox{90}{\phantom{aa} $\rho_\mathrm{core} = 10 \rho_{{\rm core},\odot}$}
    \includegraphics[height=\templengthbla cm]{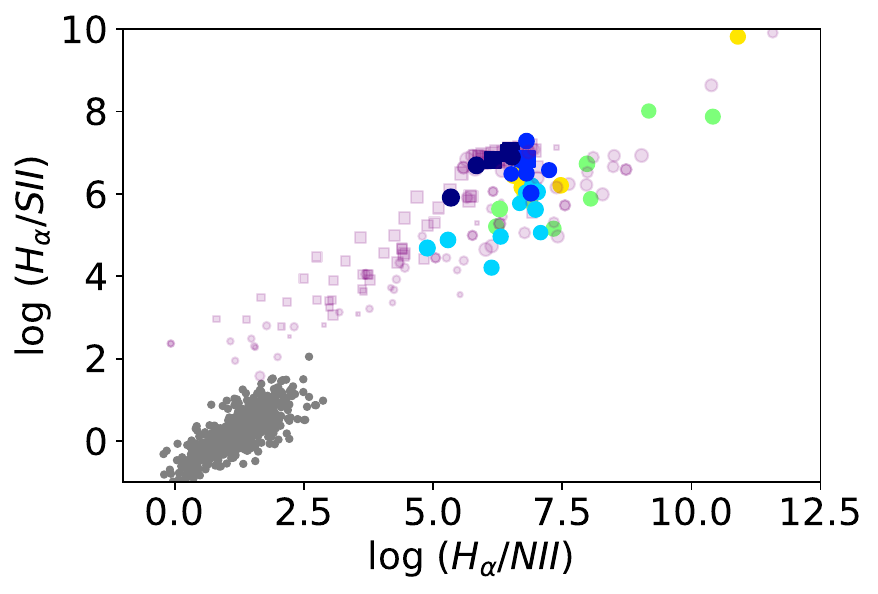}
    
    &
    \hspace*{-6mm}
    \includegraphics[height=\templengthbla cm]{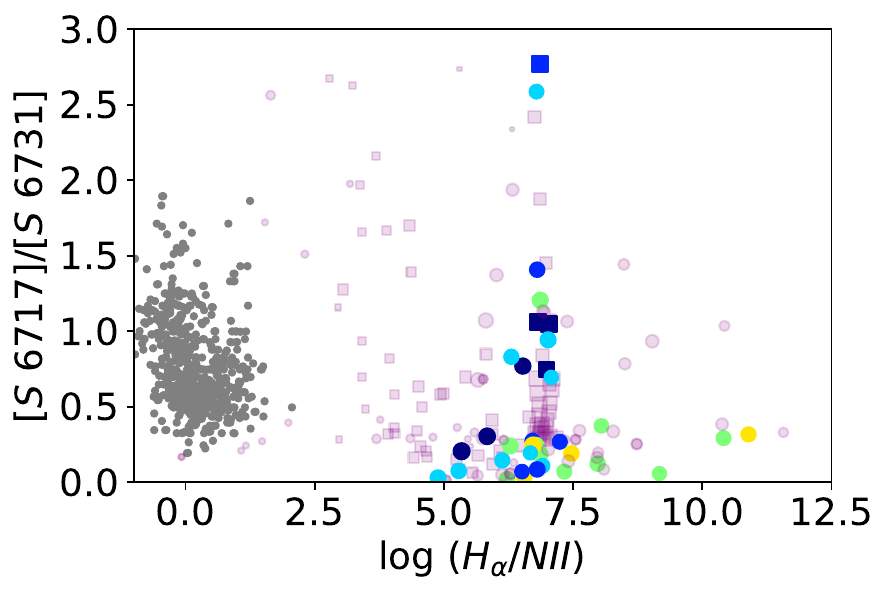}

     \\
     \rotatebox{90}{\phantom{a} $\rho_\mathrm{core} = 100 \rho_{{\rm core},\odot}$}
    \includegraphics[height=\templengthbla cm]{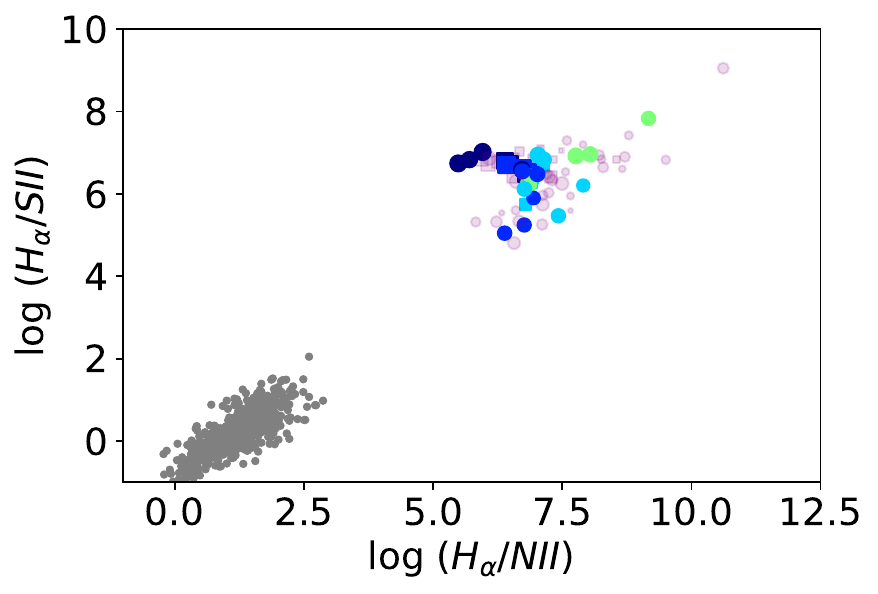}
    
    &
    \hspace*{-6mm}
    \includegraphics[height=\templengthbla cm]{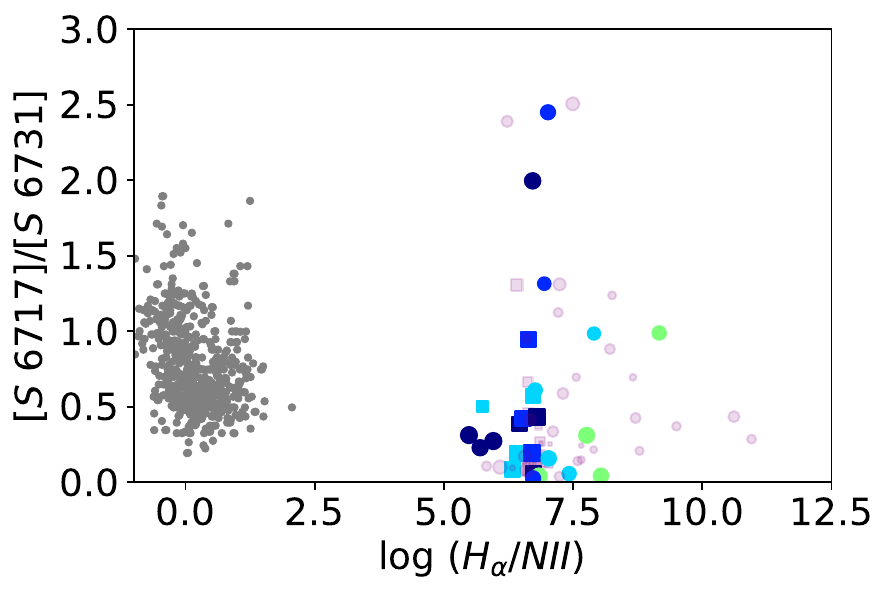}
    
    \end{tabular}
    \caption{
    Correlation between the H$\alpha$/N\,II and H$\alpha$/S\,II ratios on the left, and between H$\alpha$/N\,II and H$\alpha$ and $[S 6717]/[S 6731]$ on the right, for mirror star nuggets (colored markers) and planetary nebulae (data from \cite{Riesgo_PNpaper}), for different $\rho_{{\rm core}}$ (rows). 
    }
    \label{f.lineratios}
\end{figure}

In \fref{f.spectra}, we show a few representative white dwarf spectra from the public SDSS (BOSS) data set \citep{WD_spectra_paper} with spectral classes DA and DB. White dwarfs showing only hydrogen absorption lines are classified as spectral type DA, and are the most common spectral class. The next most common white dwarf spectra class is DB, which are characterized by He\,I lines \citep[e.g.,][]{WD_spectra_paper}.
The WD spectra are compared to  nugget spectra for several $(\rho_{{\rm core}}, M_{{\rm nugget}}, \xi)$ values. 
The two types of objects are easily distinguished by their spectra, as WDs have hot blackbody continua with pressure-broadened absorption lines, whereas nugget spectra show significant emission lines on a bright nebular continuum.
Any spectral measurement of a mirror star's optically thin nugget would conclusively demonstrate that this is no standard star.

On the other hand, the emission of optically thin nuggets is similar in character to that of planetary nebulae, with collisionally-excited and recombination lines superimposed on a nebular continuum. 
Planetary nebulae are routinely characterized by the relative strengths of specific emission lines \citep{ Baldwin1981_BPT,Riesgo_PNpaper}, which  allow the 
physical conditions within the nebula (such as density, electron temperature, and abundances),
to be determined.
Our nuggets are distinguished from planetary nebulae by their high densities, heating and ionization by collisions rather than by photo-ionization or the photoelectric effect, and the absence of any blended or scattered stellar continuum. These properties influence the Balmer and Paschen jumps, as well as lines of critical density below that of the nebula \citep{zhang2004electron,Guo2022OnContinuum}.
We expect our mirror star nuggets to exhibit different line ratios from planetary nebulae, and especially to exhibit relatively bright H lines because these are less affected by collisional de-excitation (i.e., their critical densities are very high) as well as a lack of fluorescence lines and lines from ions with super-thermal ionization potentials. These expectations is borne out in \fref{f.lineratios}, where we show the correlation between the H$\alpha$/N\,II and H$\alpha$/S\,II ratios on the left, and between H$\alpha$/N\,II and H$\alpha$ and [S\,6717]/[S\,6731] on the right, for mirror star nuggets (colored markers) and planetary nebulae (data from \citealt{Riesgo_PNpaper}). 
Again, nugget marker color and size shows heating rate and nugget mass, but unobservable nuggets ($M_G > 30$) are separately shown as purple markers to demonstrate the continuous nature of the nugget distribution in this parameter space.
As expected, the H$\alpha$/metal line ratios are much higher 
for observable nuggets than for planetary nebulae owing to the former's much higher density.

\section{Conclusions}
\label{s.conclusion}

Mirror stars are a spectacular signature of dissipative dark matter models like atomic dark matter and twin baryons in the Mirror Twin Higgs. Their discovery would provide unambiguous evidence of physics beyond the Standard Model, and reveal detailed information about the nature of the dark sector. 
Mirror Stars can emit faint electromagnetic signals through the capture, heating and excitation of SM matter from the interstellar medium.
This would allow telescopes to discover mirror stars by direct observation, but this requires a realistic and detailed prediction of their emission spectrum.

In this work, we supply comprehensive predictions for the emissions of optically thin mirror star nuggets. 
This not only defines a remarkably narrow mirror star signal region on the HR diagram (\fref{f.HR}), it also informs purely spectral methods of searching for mirror stars and distinguishing them from standard astrophysical point sources. Optically thin nuggets have a continuum spectrum that is very different from the black-body like emissions of white dwarfs and other stars; while they can be distinguished from planetary nebulae and other optically thin sources by their  emission line ratios and continuum features, indicative of their much higher density than other optically thin gas accumulations at $T \sim 10^4$\,K. 

Future work will analyze the properties and emissions of optically thick nuggets, with accumulated masses and mirror star core densities higher than the ranges studied here. This will ultimately lead to a comprehensive understanding of mirror star optical and X-ray signals.

Our results enable a variety of  new searches for mirror stars, in astrometry data, spectral surveys, and other observational catalogues. 
Dark matter could therefore be directly observed in telescopes, opening up exciting new discovery opportunities. 

\section{Acknowledgments}
We thank Peter van Hoof for conversations regarding \Cloudy.

The work of IA, BG and DC was in part supported by Discovery Grants from the Natural Sciences
and Engineering Research Council of Canada, the
Canada Research Chair program, 
the Ontario Early Researcher Award, and the University of Toronto McLean Award. BG was additionally supported by the University of Toronto Excellence Award.
The work of CM was supported by an NSERC Discovery Grant.
This research was enabled by computing resources and support provided the Digital Research Alliance of Canada (alliancecan.ca).

\bibliography{ref}

\end{document}